\documentclass[aip,jcp,reprint,showpacs,showkeys,superscriptaddress,longbibliography]{revtex4-1} % nobalancelastpage, endfloats*, linenumbers
\usepackage[utf8]{inputenc}
\usepackage[UKenglish]{babel}

\usepackage{cmap} % make ligatures etc. copy- and searchable
\usepackage[T1]{fontenc}
\usepackage{microtype}
\usepackage{amsmath,mathtools,mleftright}

\makeatletter\let\alloc@latex\alloc@%
\def\alloc@#1#2#3#4{\newcount}% the current \newcount doesn't use \alloc@

\IfFileExists{mtpro2.sty}
{ % use MathTime Professional 2
  \usepackage{times}
  \usepackage[mtphrb,mtpscr,subscriptcorrection,slantedGreek,zswash]{mtpro2} % mtpcal
}
{ % % searching the PDF for ligatures like in "diffusion" does not work with the
  \usepackage{txfonts,amssymb}
  
  \usepackage[scr=boondoxo]{mathalfa} % \mathscr used for calligraphic symbol 'N'
  \let\wtilde\widetilde
  \let\what\widehat
}
\let\alloc@\alloc@latex\makeatother

\usepackage[scr=false]{rsfso} % calligraphic math font Formal Script, likely to be used by J Chem Phys

\usepackage{graphicx,grffile,textcomp}
\graphicspath{{figures/}}
\newlength{\figwidth}

\usepackage{enumitem}
\usepackage{siunitx}
\usepackage{booktabs,tabularx,dcolumn}
\newcolumntype{d}[1]{D{.}{.}{#1}}

\usepackage{url,hyperref}
\usepackage[usenames,dvipsnames]{xcolor}
\hypersetup{colorlinks=true, linkcolor=black, %BrickRed,
urlcolor=blue!50!black, citecolor=blue!50!black}

\usepackage[capitalize]{cleveref}
\crefrangelabelformat{equation}{(#3#1#4)$-$(#5#2#6)}
\crefname{section}{Sect.}{Sects.}

\renewcommand\epsilon{\varepsilon}
\renewcommand\phi{\varphi}
\renewcommand\theta{\vartheta}
\renewcommand\rho{\varrho}
\renewcommand\leq{\leqslant}
\renewcommand\geq{\geqslant}
\renewcommand\vec[1]{\textrm{\bfseries #1}}
\newcommand\diff{\mathrm{d}}
\newcommand\expect[1]{\mleft\langle{#1}\mright\rangle}
\newcommand\e{\text{e}}
\renewcommand\i{\text{i}}

\newcommand\kB{k_{\text{B}}}
\begin{document}

\figwidth=3.4in % about \linewidth (=246.0pt)

\title{Structure of liquid--vapor interfaces: perspectives from liquid state theory, large-scale simulations, and potential grazing-incidence X-ray diffraction}

\author{F. Höf{}ling}
\email{f.hoefling@fu-berlin.de}
\affiliation{Freie Universität Berlin, Fachbereich Mathematik und Informatik,
Arnimallee 6, 14195 Berlin, Germany}
\affiliation{Zuse Institut Berlin, Takustr. 7, 14195 Berlin, Germany}

\author{S. Dietrich}
\affiliation{Max-Planck-Institut für Intelligente Systeme, Heisenbergstraße 3,
70569 Stuttgart, Germany}
\affiliation{IV. Institut für Theoretische Physik,
Universität Stuttgart, Pfaffenwaldring 57, 70569 Stuttgart, Germany}

\date{\today}

\begin{abstract}
\noindent
Grazing-incidence {X}-ray diffraction (GIXRD) is a scattering technique which allows one to characterize the structure of fluid interfaces down to the molecular scale, including
the measurement of the surface tension and of the interface roughness.
However, the corresponding standard data analysis at non-zero wave numbers has been criticized
as to be inconclusive because the scattering intensity is polluted by the unavoidable
scattering from the bulk.
Here we overcome this ambiguity by proposing a physically consistent model
of the bulk contribution which is based on a minimal set of assumptions of experimental relevance.
To this end, we derive an explicit integral expression for the background scattering,
which can be determined numerically from the static structure factors of the coexisting bulk phases as independent input.
Concerning the interpretation of GIXRD data inferred from computer simulations, we extend the model to account also for the finite sizes of the bulk phases, which are unavoidable in simulations.
The corresponding leading-order correction beyond the dominant contribution to the scattered intensity is revealed by asymptotic analysis,
which is characterized by the competition between the linear system size and the {X}-ray penetration depth in the case of simulations.
Specifically, we have calculated the expected GIXRD intensity for scattering at the planar
liquid--vapor interface of Lennard-Jones (LJ) fluids with truncated pair interactions via extensive, high-precision computer simulations.
The reported data cover interfacial and bulk properties of fluid states along the whole liquid--vapor coexistence line.
A sensitivity analysis shows that our findings are robust with respect to the detailed definition of the mean interface position.
We conclude that previous claims of an enhanced surface tension at mesoscopic scales are amenable to unambiguous tests via scattering experiments.
\end{abstract}

% \pacs{68.03.Cd, 61.20.Ja, 05.10.-a}

% 05.10.-a    Computational methods in statistical physics
% 68.03.Cd    Surface tension and related phenomena
% 61.20.Ja    Computer simulation of liquid structure
% 61.25.Em    Molecular liquids
% 61.20.Ne    Structure of simple liquids
% 68.03.Hj    Liquid surface structure: measurements and simulations

\maketitle

\section{Introduction}

The liquid--vapor interface is a ubiquitous confining boundary of fluids and has been the subject of enduring experimental, theoretical, and simulational interest. These efforts focus on properties of
adsorbed liquid films \cite{Fernandez:2012,Pottier:2014,MacDowell:2014,MacDowell:2017},
droplets \cite{Aarts:2004, Horsch:2012, Malijevsky:2012, Bruot:2016, Troester:JPCB2017, Aasen:2018},
and interfaces out of equilibrium \cite{Jackson:2016,Heinen:2016,Midya:2017,Muscatello:2017,delJunco:2019,Zhang:L2021},
as well as on applications such as
wetting \cite{Rauscher:ARMR2008,Bonn:2009,Evans:2017,Pang:2019},
solvation \cite{Blanc:2013,Rane:2016,Rane:2016a,Bickel:2014,Hernandez-Munoz:PRB2019},
and the presence of surfactants \cite{Calzolari:SM2012,Costa:NL2016,Dasgupta:JPCM2017,Scoppola:COCIS2018,Jaksch:COCIS2019,Zhang:PRF2023,Ukleev:2024}.
A considerable number of investigations is motivated by the long-standing issue concerning the influence of van der Waals or dispersion forces on the structure and behavior of the interface \cite{Napiorkowski:1993, Parry:1994, Mecke:1999, Fradin:2000, Mora:2003, Li:2004, Mueller:2000, Milchev:2002, Chacon:2003, Vink:2005, Blokhuis:2008, Blokhuis:2009, Sedlmeier:2009, Chacon:2014, Nagata:2016, Sega:2017}.
Not only due to this issue, the liquid--vapor interface has been an important testing ground for the theory of inhomogeneous fluids.
Recent advances concerning the theoretical description have been stimulated by simulation data for a range of temperatures \cite{Capillary:2015}
and led to accurate predictions of interfacial density correlations by combining the concept of a position-dependent, local structure factor \cite{Parry:2014, Parry:2015, Parry:2016}
and insight into resonances stemming from the bulk structure \cite{Parry:NP2019, Parry:PRE2019, Parry:PRE2019b}.
In view of these predictions, a fully comprehensive interpretation of experiments concerning liquid--vapor interfaces still remains to be formulated.

Experimentally, scattering techniques such as X-ray reflectometry and grazing-incidence X-ray diffraction (GIXRD) provide the most detailed view of fluid interfaces. Whereas reflectometry allows one to infer interfacial density profiles,
GIXRD probes density fluctuations \cite{Dietrich:1995,Daillant:2000,Pershan:LiquidSurfaces} and surface structures \cite{Scoppola:COCIS2018,Jaksch:COCIS2019}.
This kind of scattering experiments under grazing incidence have been carried out for liquid surfaces of water \cite{Fradin:2000}, molecular fluids \cite{Mora:2003}, and liquid metals \cite{Li:2004}.
The analysis of the scattered intensities requires accounting for the inevitable scattering from the fluid bulk phases, which depends on a suitable background model \cite{Mora:2002,Paulus:2008,Pershan:2009},
nowadays available within GIXRD analysis software \cite{Pospelov:JAC2020};
yet, such a modeling choice can potentially introduce ambiguities into the interpretation of the experimental data.

Focusing on thermal equilibria of coexisting liquid and vapor phases,
the increase of the interfacial area by a macroscopic amount $\Delta A$ incurs a free energy cost of $\gamma_0 \, \Delta A$,
which defines the (macroscopic) surface tension $\gamma_0$.
At the molecular scale, the interface is roughened by thermal fluctuations, which are tamed by an analogous cost in free energy.
This mesoscopic picture leads to the capillary wave (CW) theory \cite{Buff:1965,Evans:1979,Rowlinson:Capillarity},
which assumes the interface to be a smooth, membrane-like surface with a tension modulus which is governed by an interface Hamiltonian.
On the other hand, by adopting the particle perspective, the theory of inhomogeneous fluids\cite{Evans:1979, Henderson:InhomogeneousFluids, Hansen:SimpleLiquids} considers (two-point) density fluctuations in order to characterize the interfacial region.
For a free, planar interface, the classical result by Wertheim \cite{Wertheim:1976} and Weeks \cite{Weeks:1977} states that fluctuations with wavevector $\vec q$ parallel to the surface
lead to a scattering intensity proportional to $\kB T / (\gamma_0 q^2)$, which applies for small $q=|\vec q|$, i.e., for macroscopic wavelengths $2\pi/q$ [see, cf., \cref{fig:gid_analysis,eq:gamma_q_ell}]; % \cref{sec:interfaces}
the thermal energy scale is denoted by $\kB T$ as usual.
This asymptotic result $(q \to 0)$ matches with the free energy cost of the excitation of corresponding CWs. It has stimulated elaborate derivations of effective interfacial Hamiltonians \cite{Napiorkowski:1993, Parry:1994, Mecke:1999, Hiester:2006, Chacon:2016, Hernandez-Munoz:2016, *Hernandez-Munoz:2018, *Hernandez-Munoz:2018a}
in order to capture also the (anticipated) structure of the two-point correlation function at higher orders in~$q$.
These extended CW theories include also a bending modulus $\kappa$ of the interface \cite{Blokhuis:2008, Blokhuis:2009} or, more generally, a wave number-dependent surface tension $\hat\gamma(q)$ which exhibits the relation $\hat\gamma(q\to 0) = \gamma_0$.
The leading correction to this macroscopic limit is either non-analytic (of the form $q^2 \log(q)$) or quadratic ($\kappa q^2$), depending on whether dispersion forces are present or not.
Molecular dynamics (MD) simulations testing the behavior of $\hat\gamma(q)$ have frequently been analyzed in line with the CW picture by evaluating the fluctuation spectrum of the local interface position \cite{Milchev:2002, Vink:2005, Blokhuis:2008, Blokhuis:2009, Sedlmeier:2009, Pang:2019, Chacon:2003, Bresme:2008a, Tarazona:2012, Chacon:2014, Hernandez-Munoz:JCP2022}.
This requires one to adopt a suitable definition of this position down to the molecular scale.
Common choices are either a local Gibbs dividing surface or an elaborate algorithm to identify an intrinsic interface which separates CWs from bulk-like density fluctuations.
However, in view of the various possible choices the published data for $\hat\gamma(q)$ disagree on even the sign of the bending coefficient $\kappa$ (in the absence of dispersion forces).
Moreover, it was shown within density functional theory (DFT) \cite{Parry:2014} that any effective interface Hamiltonian fails to reproduce the behavior of interfacial density correlations, which are obtained from simulations.

\begin{figure}
  \fboxsep=0pt
  \fbox{\includegraphics[width=\figwidth]{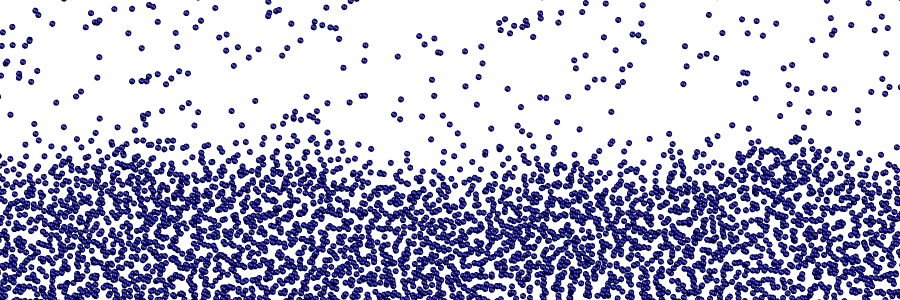}} \\[1em]
  \fbox{\includegraphics[width=\figwidth]{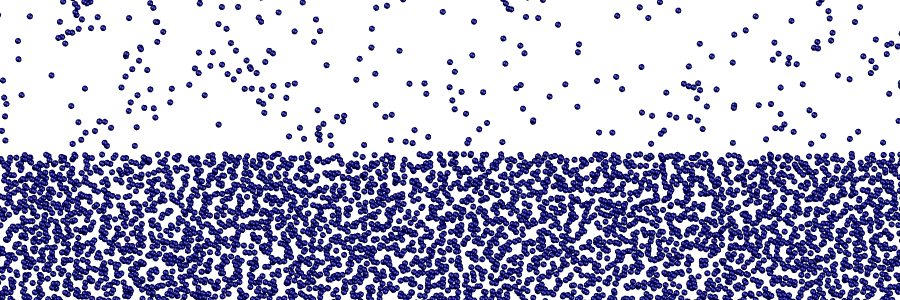}}
  \caption{Top: cross-section of a snapshot of a simulated liquid--vapor interface configuration for a truncated Lennard-Jones (LJ) fluid at liquid--vapor bulk coexistence for the temperature $T \approx 0.8 T_c$; $T_c$ is the critical temperature of this fluid.
  Bottom: interface-related fluctuations are switched off in a \emph{Gedankenexperiment} (see the main text). This snapshot depicts the initial configuration of the simulation, composed of two independent bulk configurations, before further equilibration.}
  \label{fig:interface2d}
\end{figure}

As an alternative which bypasses the ambiguities associated with the definition of a local interface position, we previously introduced an effective surface tension $\gamma(q)$ as a proxy for interfacial density fluctuations which is entirely based on quantities amenable to scattering experiments \cite{Capillary:2015} [see below, \cref{eq:gamma_q_def}].
Similarly, as for the interpretation of scattering data, this approach hinges on a consistent model for the background scattering, which is the subject of the present study.
It is based on a hypothetical liquid--vapor interface with \emph{all} interface-related correlations switched off (\cref{fig:interface2d}), for which the scattered intensity can be worked out analytically.
Somewhat unexpectedly, these open boundary conditions at the interface lead to corrections in the small-wave number density correlations due to bulk fluctuations \cite{SSFslab:JCP2020}, which, as it will turn out, interfere with the small-$q$ behavior of $
\gamma(q)$; a related boundary effect has been observed for the two-dimensional strip geometry \cite{Delfino:JHEP2016,Squarcini:JSM2021}.
Accounting for these corrections removes an inconsistency between the results for $\gamma(q)$ obtained from the direct calculation of the density correlations and obtained from simulated scattering intensities.
Moreover, it renders the bending coefficient in $\gamma(q)$ positive for all temperatures at liquid--vapor coexistence, between the triple point and the critical point of the fluid.

The paper is organized as follows. In \cref{sec:preliminaries} we provide a number of relations which will be useful for the theoretical analysis; in particular, we discuss the separation of the GIXRD intensity into interfacial and background parts, provide the definition of $\gamma(q)$, and connect with Ref.~\citenum{SSFslab:JCP2020}.
The background scattering for the above mentioned reference system is analyzed in \cref{sec:reference} with special emphasis on the non-commuting limits of infinite X-ray penetration depth and infinite sample width.
In \cref{sec:interfaces}, these results are applied to the analysis of simulation data for scattering intensities from liquid--vapor interfaces.

\section{General considerations}
\label{sec:preliminaries}

\subsection{GIXRD master formula}

For diffraction experiments on planar interfaces under grazing incidence,
the scattering intensity is proportional to
the two-point correlation function of the atomic number density.
The dependence of the scattering intensity on the lateral scattering vector $\vec q=(q_x, q_y)$
follows the master formula
[Eq.~(2.68) in Ref.~\onlinecite{Dietrich:1995}]
\begin{equation}
  I(q) = \iint\limits_{\mathbb{R}\times \mathbb{R}} \! \diff z \,\diff z' \, f(z)^* f(z') \, G(q, z, z')
  \label{eq:gid_master}
\end{equation}
with $q=|\vec q|$ and the $z$-axis coinciding with the interface normal.
Here, we have dropped the atomic form factors for reasons of simplicity (i.e., implicitly considering point-like particles)
and have omitted the reflection and transmission coefficients, which are constant amplitudes with respect to the lateral momentum transfer.\footnote{On the vapor side, the integral in \cref{eq:gid_master} is divergent. However, this issue is neither present in the simulations, due to the finite size of the simulation box, nor in experiments, due to the (weak) absorption of X-rays in the vapor phase.}
The two-point number density correlation function $G(q, z, z')$ characterizes the
sample and is independent of the experimental setup probing it.

Within the theory of fluids, this density--density correlation function is defined as \cite{Evans:1979,
Hansen:SimpleLiquids}
$G(\vec r, \vec r')=\expect{\hat\rho(\vec r) \, \hat\rho(\vec r')}
  - \expect{\hat\rho(\vec r)} \expect{\hat\rho(\vec r')}$,
where $\hat\rho(\vec r)$ is the microscopic number density and
$\rho(\vec r) = \expect{\hat\rho(\vec r)}$ is the mean density at point $\vec r = (\vec R, z)$.
It is convenient to split off the singular self-part of $G(\vec r, \vec r')$
and to introduce the pair correlation function $g(\vec r, \vec r')$ by
\begin{equation}
  G(\vec r, \vec r') = \rho(\vec r)\, \rho(\vec r') \, [g(\vec r, \vec r') - 1] + \rho(\vec r) \,\delta(\vec r - \vec r') \,.
  \label{eq:G_general}
\end{equation}
In planar geometry, translational invariance parallel to the mean interface entails
$\rho(\vec r) = \rho(z)$ and $G(\vec r, \vec r') = G(\vec R - \vec R', z, z')$.
This suggests to consider the lateral Fourier transform.  Accordingly, for the two-component vectors
$\vec q$ and $\Delta\vec R = \vec R - \vec R'$, one introduces
\begin{align}
  G(|\vec q|, z, z')
    &:= \int_{\mathbb{R}^2} \diff^2 (\Delta R) \, \e^{-\i \vec q \cdot \Delta \vec R} \, G(\Delta \vec R, z, z') \,,
  \label{eq:Gq_def}
\end{align}
which enters into \cref{eq:gid_master}.
$G(q, z, z')$ depends only on the modulus $q=|\vec q|$ of the wavevector due to the isotropy of the sample in the planes parallel to the interface.

The function $f(z)$ is determined by the scattering geometry (i.e., the angles
$\alpha_i$ and $\alpha_f$ of incoming and outgoing beams, respectively; see Fig.~2 in Ref.~\onlinecite{Dietrich:1995})
and the mean density profile $\rho(z)$.
It describes the decay of the evanescent wave on the liquid side of the
interface, $f(z<0) \sim \exp(-\kappa |z|)$, with the penetration depth
$1/\kappa(\alpha_i,\alpha_f)$ \cite{Dietrich:1984, Dietrich:1995}.
For $\alpha_i$ and $\alpha_f$ approaching the critical angle of total
reflection, the penetration depth diverges, and, in the thermodynamic limit,
$I(q)$ is dominated by scattering from the liquid bulk.
Thus, experimental setups aim at minimizing the penetration depth $1/\kappa$,
which in practice can reach down to a few nanometres \cite{Fradin:2000,Mora:2003}
or even below \cite{Murphy:N2016,Konovalov:2022}.
On the other hand, one has to ensure that the penetration depth $1/\kappa$ is considerably larger than the interfacial width $\zeta$ (cf.~\cref{eq:sigmoidal_profile});
otherwise, the observation of interfacial fluctuations would be incomplete.
Therefore, the condition to access interfacial properties fully is $\kappa \zeta \ll 1$, which is met by the typical experimental setups; in analytical and simulation work, it implies that the undamped limit, $\kappa \to 0$, is considered.

\subsection{Capillary wave divergence}

Concerning the $q$-dependence of the scattering intensity,
classical capillary wave theory predicts, for the density correlations at small wave numbers, that asymptotically \cite{Wertheim:1976,Weeks:1977,Evans:1979}
\begin{equation}
  G(q, z, z') \simeq \kB T \, \frac{\rho'(z) \, \rho'(z')}{\gamma_0 \, (q^2 + \ell_c^{-2})}\,, \qquad q \to 0,
  \label{eq:WertheimWeeks}
\end{equation}
where $\gamma_0$ denotes the macroscopic surface tension and $\ell_c = \sqrt{\gamma_0/(m g\Delta \rho)}$ is the capillary length;
$\Delta \rho = \rho_\ell - \rho_v$ is the number density contrast between the coexisting phases, and $m g$ is the gravitational weight of a molecule of mass $m$.
Clearly, $G(q, z, z')$ diverges as $q^{-2}$ for large capillary lengths, $\ell_c \gg q^{-1}$, which is considered in the following.
In combination with \cref{eq:gid_master}, the divergence is passed on to the scattered intensity:
\begin{equation}
  I_\text{CWT}(q \to 0) \simeq \frac{\kB T (\Delta \rho)^2}{\gamma_0 \, q^2} .
  \label{eq:I_CWT}
\end{equation}
Since the bulk scattering remains finite as $q\to 0$, the interfacial contribution dominates the signal for small~$q$.
In order to obtain information about the interface which goes beyond this divergence, one needs to unambiguously separate interface-related and bulk contributions to the scattering intensity:
\begin{equation}
  I(q;\kappa) = I_\text{int}(q;\kappa) + I_b(q;\kappa) \,.
  \label{eq:Iq_sep}
\end{equation}
This leads one to introduce the interfacial structure factor as
\begin{equation}
  H(q) := \lim_{\kappa  \to 0} [I(q;\kappa) - I_b(q;\kappa)] = I_\text{int}(q;\kappa = 0)
  \label{eq:H_q}
\end{equation}
and an effective, wave number-dependent surface tension $\gamma(q)$ by generalizing \cref{eq:I_CWT}:
\begin{equation}
 H(q) =: \frac{\kB T (\Delta \rho)^2}{\gamma(q) \, q^2} \,.
 \label{eq:gamma_q_def}
\end{equation}
The macroscopic surface tension is recovered as $\gamma_0 = \gamma(q \to 0)$ and the $q$-dependence of $\gamma(q)$ quantifies the deviations from the divergence $H(q \to 0) \sim q^{-2}$, which can be attributed to interface-related density fluctuations.

The division in \cref{eq:Iq_sep} hinges on a consistent model for the bulk scattering $I_b(q;\kappa)$ which is based on clearly formulated assumptions and on experimentally accessible quantities only.
A simple and commonly used model is based on the bulk structure factor of the liquid phase \cite{Fradin:2000,Mora:2003}:
\begin{equation}
  I_b(q;\kappa) \approx \frac{\rho_\ell S_\ell(q)}{2\kappa} \simeq \frac{\rho_\ell^2 \, \kB T \chi_T}{2\kappa}  \,, \quad q \to 0,
\end{equation}
where $\chi_T$ is the isothermal compressibility.
We shall show below that this approximation, even in the small-$q$ regime, creates a constant shift in the interfacial structure factor $H(q)$ and thus a bias of $O(q^2)$ in $\gamma(q)$.
Moreover, it leads to inconsistencies between experiments and simulations.

\subsection{Structure factor of an open slab of liquid}
\label{sec:ssf_slab}

In Ref.~\onlinecite{SSFslab:JCP2020}, we investigated
the effect of open boundaries of a liquid sample of finite width $L< \infty$.
It revealed a finite-size correction to the structure factor and anticipates the solution strategy followed in \cref{sec:reference}.
This study is based on virtually excavating a planar slab of width $L$ from a homogeneous liquid, thereby imposing free boundary conditions on the newly created surfaces (see Fig.~1 of Ref.~\onlinecite{SSFslab:JCP2020}; the corresponding situation with equilibrated interfaces is depicted below in \cref{fig:snapshot-3d}). The observable of interest is the slab structure factor,
\begin{equation}
  S(q;L) = \frac{1}{\rho L} \iint\limits_{\mathclap{0 \leq z, z' \leq L}} \diff z\,\diff z' \, G(q, z, z'),
  \label{eq:gid_master_slab}
\end{equation}
which, up to the prefactor, resembles \cref{eq:gid_master} for a step function $f(z) = 1_{[0,L]}(z)$, chosen as the indicator function of the interval $[0,L]$; the latter emerges in the limit $\kappa \to 0$ and for a finite system.
Using the two-point correlation function of a homogeneous liquid, $G(q, z, z') = G_\ell(q, \Delta z = z - z')$, yields
$\rho L S(q; L) = I_\ell(q; \kappa \to 0, L)$ [see \cref{eq:gid_master,eq:Iq_sep}].
The crucial step in the derivation is the observation that for a homogeneous fluid, the pair correlation function is fully determined by the bulk structure factor. This is expressed by the relation
\begin{equation}
  G_\ell(|\vec q|, \Delta z)
  = \rho_\ell \! \int \!\frac{\diff k_z}{2\pi} \,\e^{\i k_z \Delta z} \, S_\ell(|\vec k|)
      - (2\pi \rho)^2\delta(\vec q) \,,
  \label{eq:Gq_bulk}
\end{equation}
which is obtained by a partial Fourier back transform with $\vec k = (-\vec q, k_z)$ denoting a three-component wave vector.
The subscripts $\ell$ indicate that $\rho_\ell$ and $S_\ell$ refer to the
number density and the structure factor of the bulk liquid, respectively.
Inserting $G_\ell(|\vec q|, \Delta z)$ into \cref{eq:gid_master_slab} and carrying out the integrals over $z$ and $z'$ yields \cite{SSFslab:JCP2020}
\begin{multline}
  S(|\vec q|; L)
    = \frac{2}{\pi L} \! \int_0^\infty \!\diff k_z \,
      \frac{1-\cos(k_z L)}{k_z^2} \,
      S_\ell\Bigl(\!\sqrt{|\vec q|^2 + k_z^2}\Bigr) \\
      + \rho L (2\pi)^2 \delta(\vec q) \,.
  \label{eq:ssf_slab}
\end{multline}
For wide slabs, one finds the asymptotic expansion
\begin{equation}
  S(q>0;L \to \infty) = S_\ell(q)  + 2 L^{-1} \mathcal{J}_0(q) + O\bigl(L^{-1}\e^{-L/\xi}\bigr) \,,
  \label{eq:ssf_slab_asymptotics}
\end{equation}
where $\mathcal{J}_0(q)$ is the leading-order correction integral
\begin{equation}
  \mathcal{J}_0(q) := \frac{1}{\pi} \int_0^\infty\! \diff k_z \,
    \frac{S_\ell\bigl(\sqrt{q^2 + k_z^2}\bigr) - S_\ell(q)}{k_z^2} \,.
  \label{eq:correction_integral}
\end{equation}
Most importantly, the small-wave number limit $\mathcal{J}_0(q \to 0)$ is non-zero and can have either sign, depending on, e.g., the temperature (see below, \cref{fig:correction_integral}).
We stress that this kind of finite-size corrections does not appear for periodic boundary conditions at the surfaces $z=0$ and $z=L$ (as commonly used in simulations). In the latter case, the equation $S_\text{per}(q;L) = S_\ell(q)$ holds exactly.

\section{Bulk reference with open boundaries}
\label{sec:reference}

\subsection{Density--density correlation function}
\label{sec:Gb_model}

The interface-related contributions to the scattering intensity are unambiguously identified via the comparison with a background reference system in which all interface-induced perturbations are switched off. This idealized situation can be
created by constructing an ensemble of particle configurations, which contain a planar interface and which are characterized solely by bulk correlations. Such configurations are obtained
in a \emph{Gedankenexperiment} by virtually cutting a homogeneous liquid sample of macroscopic size along a plane, denoted as $z=0$, and by removing all molecules above the plane (i.e., with positions $z>0$). The empty half space is then filled by a correspondingly treated sample of the coexisting vapor phase.
This renders a flat liquid--vapor interface without any structural distortions in the vicinity of the plane $z=0$ (\cref{fig:interface2d}). By construction, the local densities on opposite sides of the interface are independent of each other and, in particular, uncorrelated.
On the other hand, for $z$ and $z'$ both being on the same side, the two-point correlation function of the reference system equals that of the respective bulk phase, no matter how close to the interface $z$ and $z'$ are chosen to be.
Thus, we define the two-point correlation function of this ``naked interface'' ensemble, which serves as background reference, as
\begin{equation}
  G_b(q,z,z') =
  \begin{cases}
    G_\ell(q, z - z'), & z, z' < 0, \\
    G_v(q, z - z'), & z, z' > 0, \\
    0, & z \cdot z' \leq 0.
  \end{cases}
  \label{eq:Gb_model}
\end{equation}
It describes an inhomogeneous system of two unperturbed and uncorrelated, coexisting bulk phases which share a flat, open boundary; due to the spatial homogeneity of the bulk phases, $G_\ell$ and $G_v$ depend only on $z-z'$ instead on $z$ and $z'$ separately.
It is clear that the dynamic evolution of these reference configurations would necessarily lead to the usual equilibrium with fully developed interfacial fluctuations; however, statistical averages of the background configurations are understood as ensemble averages, not time averages.

The bulk scattering follows by inserting this definition of a reference system [i.e., \cref{eq:Gb_model}] into the master formula, [\cref{eq:gid_master}] which is carried out in \cref{sec:half_space}.
Concerning simulations of GIXRD experiments, the same model for the bulk correlations can be used with the only modification that the non-zero width of the liquid film must be taken into account. Accordingly, the scattering depends on both the penetration depth $1/\kappa$ and the sample width $L$ with the relevant, non-commuting limits $\kappa \to 0$ and $L \to \infty$ (see \cref{sec:finite_slab}).

We note that $G_b(q,z,z')$ as defined in \cref{eq:Gb_model} is discontinuous at the interface as the reference system changes abruptly between the two phases.
On the other hand, the physically observed, full two-point correlation $G(q,z,z')$ is a continuous function of $z$ and $z'$ for all wave numbers $q$.
Any smooth interpolation of the background between the coexisting phases would necessarily make assumptions on the correlations in the interfacial region, e.g., it would introduce an unknown interpolation length.
\footnote{Such interpolation schemes are possible for exactly solvable microscopic models \cite{Parry:2015}.}
A discontinuity of $G_b(q, z, z')$ implies that the interfacial part $G_\text{int}(q,z,z') := G(q, z, z') - G_b(q, z, z')$ is also discontinuous.
We emphasize that this conceptually unavoidable discontinuity of $G_b$ and thus $G_\text{int}$ is not in conflict with theoretical constraints \cite{Parry:2015}; most importantly, it does not contradict the asymptotically rigorous result for the CW divergence [\cref{eq:WertheimWeeks} with $l_c = 0$]:
\begin{equation}
  \lim_{q\to 0} q^2 G(q,z,z') = (\kB T/\gamma_0) \rho'(z) \rho'(z') \,,
  \label{eq:CW-divergence}
\end{equation}
which is continuous in $z$ and $z'$ since the mean density profile $\rho(z)$ is a smooth function.
The same property carries over to $G_\text{int}(q,z,z')$, because $G_b(q,z,z')$ is a bounded function of $q$:
\begin{align}
  \lim_{q\to 0} q^2 G(q,z,z') &= \lim_{q\to 0} q^2 [G_\text{int}(q,z,z') + G_b(q,z,z')] \notag  \\
  &= \lim_{q\to 0} q^2 G_\text{int}(q,z,z') \,.
\end{align}
It turns out that the continuity of $G_\text{int}(q,z,z')$ is not needed for this to hold; it is sufficient that the discontinuity is uniformly bounded in $q$ (as a function of $z$ and $z'$).
The line of arguments is as follows:
Let us separate $G_\text{int} = G_\text{int}^c + G_\text{int}^\Delta$ into its continuous part $G_\text{int}^c(q,z,z')$ and its discontinuous part $G_\text{int}^\Delta(q,z,z')$, which is piecewise constant in $z$ and $z'$.
The latter is also the discontinuous part of $G_b$, and, because $G_\ell$ and $G_v$ are bounded, $G_\text{int}^\Delta(q,z,z')$ is thus uniformly bounded in $q$.
With this, in the limit $q\to 0$, the discontinuous part drops out:
\begin{align}
  \lim_{q\to 0} q^2 G_\text{int}(q,z,z') &= \lim_{q\to 0} q^2 [G_\text{int}^c(q,z,z') + G_\text{int}^\Delta(q,z,z')] \notag \\
  &= \lim_{q\to 0} q^2 G_\text{int}^c(q,z,z') \,.
\end{align}

\subsection{Scattering from a macroscopic half-space of bulk liquid}
\label{sec:half_space}

In this section, we turn to the specific setup of GIXRD experiments, i.e., a macroscopic liquid sample ($L \to \infty$) and an evanescent wave on the liquid side ($\kappa > 0$).
According to the master formula in \cref{eq:gid_master}, the GIXRD intensity for the bulk reference model [\cref{eq:Gb_model}] is the sum of two independent contributions stemming from the integrals over $G_b(q,z,z')$ in the domains $z,z' < 0$ (bulk liquid) and $z,z' > 0$ (bulk vapor).
At low vapor pressure, the scattering from the latter side of the interface is negligible; alternatively, the result of Ref.~\citenum{SSFslab:JCP2020} can be used directly because there is no damping of the propagating beam, i.e., $f(z > 0) = 1$.

The calculation of the bulk scattering from the liquid side ($z < 0$) is conceptually similar, albeit the algebraic expressions differ.
Due to the exponential decay of the evanescent wave, i.e., $f(z) = \exp(-\kappa |z|)$ for $z < 0$, the integrals in \cref{eq:gid_master} can be formulated explicitly:
\begin{align}
  I_\ell(q)
  &= \iint\limits_{z,z' < 0} \hspace{-.5em} \diff z \,\diff z' \, \e^{\kappa (z+z')} \, G_\ell(q, z - z') \notag \\
  &= \frac{1}{2}\int_{-\infty}^\infty\!\diff u \!\int_{|u|}^\infty \! \diff v\, \e^{-\kappa v} \, G_\ell(q, u) \,,
  \label{eq:gid_uv_integrals} \\
\intertext{upon a change of variables $u:=z-z'=\Delta z$ and $v:=-(z+z')$ and by taking into account the absolute value of the
Jacobian $|\partial(u,v)/\partial(z,z')|=2$.
The integral over $v$ is simple, and replacing the bulk correlation function $G_\ell(q, u)$ by
\cref{eq:Gq_bulk} yields}
  I_\ell(\vec q)
  &= \frac{1}{\kappa} \int_0^\infty\!\diff u \, \e^{-\kappa u} \,
      \biggl\{ \rho_\ell \!\int \!\frac{\diff k_z}{2\pi} \,\e^{\i k_z u} \, S_\ell\bigl(\sqrt{q^2 + k_z^2}\bigr) \notag \\
   & \hspace{10em} - \rho_\ell^2 \, (2\pi)^2\delta(\vec q) \,\biggr\} \/, \quad \kappa > 0.
\end{align}
Finally, interchanging the integrations over $u$ and $k_z$ leads to
\begin{equation}
  I_\ell(|\vec q|) = \rho_\ell \int\limits_{-\infty}^\infty \!\frac{\diff k_z}{2\pi} \,
    \frac{S_\ell\Bigl(\sqrt{|\vec q|^2 + k_z^2}\Bigr)}{\kappa^2 + k_z^2}
    - \frac{\rho_\ell^2}{\kappa^2}\, (2\pi)^2\delta(\vec q) \,.
  \label{eq:gid_bulk}
\end{equation}
For a given static structure factor $S_\ell(k)$ and wave numbers $k < k_\text{max}$, the integral can be
evaluated numerically as described in Ref.~\citenum{SSFslab:JCP2020}.
To this end, we use the identity $\int_0^\infty \bigl(1+x^2\bigr)^{-1} \,\diff x = \pi/2$
and rewrite
\begin{equation}
  I_\ell(q>0) = \frac{\rho_\ell}{\pi} \int\limits_0^\infty \!\diff k_z \,
    \frac{S_\ell\Bigl(\sqrt{q^2 + k_z^2}\Bigr) - 1}{\kappa^2 + k_z^2}
    + \frac{\rho_\ell}{2\kappa}\,,
  \label{eq:gid_bulk_numerics}
\end{equation}
which can be truncated safely at large~$k_z$, recalling that $S_\ell(k \to \infty) = 1$.
As a by-product, one finds $I_\ell(q\to  \infty) = \rho_\ell/(2\kappa)$.

In the limit $\kappa \to 0$, i.e., for scattering angles close to the angle of total reflection, one can identify a part of the integrand in \cref{eq:gid_bulk} as a representation of Dirac's $\delta$-distribution:
% \footnote{
% For every $f(x)$ integrable and continuous at $x=0$, it holds \newline
% $ \displaystyle
%   \lim_{\kappa \to 0} \frac{1}{\pi} \int_{-\infty}^\infty \frac{\kappa}{\kappa^2 + x^2} \, f(x)\,\diff x = f(0)
% $
% }
\begin{equation}
  \frac{1}{\pi} \int_0^\infty \! \phi(x)\,\frac{\kappa}{\kappa^2 + x^2} \, \diff x
  \xrightarrow{\kappa \to 0}
  \int_0^\infty \! \phi(x)\, \delta(x) \,\diff x = \frac{1}{2}\phi(0) \,,
  \label{eq:Dirac_delta}
\end{equation}
which holds for a continuous and bounded test function $\phi(x)$.
$I_\ell(|\vec q|)$ resembles
the bulk structure factor evaluated at wave vectors $\vec q$ parallel to the interface:
\begin{equation}
  I_\ell(q) \simeq \frac{\rho_\ell}{2\kappa} \,S_\ell(q) \,, \quad \kappa\to 0.
  \label{eq:gid_bulk_limit}
\end{equation}
The next-to-leading order term $O\bigl(\kappa^0\bigr)$ within the asymptotic expansion of $I_\ell(q)$ around $\kappa = 0$ is given by
\begin{align}
  \mathrlap{\lim_{\kappa \to 0} \left[ I_\ell(q) - \frac{\rho_\ell}{2\kappa} \,S_\ell(q) \right]} \qquad \notag \\
  &= \lim_{\kappa \to 0} \frac{\rho_\ell}{\pi} \int_0^\infty \! \diff k_z \,
      \frac{S_\ell\Bigl(\sqrt{q^2 + k_z^2}\Bigr) - S_\ell(q)}{\kappa^2 + k_z^2} \notag \\
  &= \rho_\ell \, \mathcal{J}_0(q) \,,
  \label{eq:gid_bulk_corr}
\end{align}
where we have taken the limit $\kappa\to 0$ inside of the integral, as permitted by the theorem on monotone convergence,
and we have made use of the definition of $\mathcal{J}_0(q)$ [\cref{eq:correction_integral}].
Combining \cref{eq:gid_bulk_limit,eq:gid_bulk_corr}, we arrive at one of our main results:
\begin{equation}
  I_\ell(q) = \frac{\rho_\ell}{2\kappa} \,S_\ell(q) + \rho_\ell \, \mathcal{J}_0(q) + O(\kappa) \,, \quad \kappa \to 0.
  \label{eq:gid_bulk_asymptotics}
\end{equation}

\begin{figure*}
  \includegraphics[width=\textwidth]{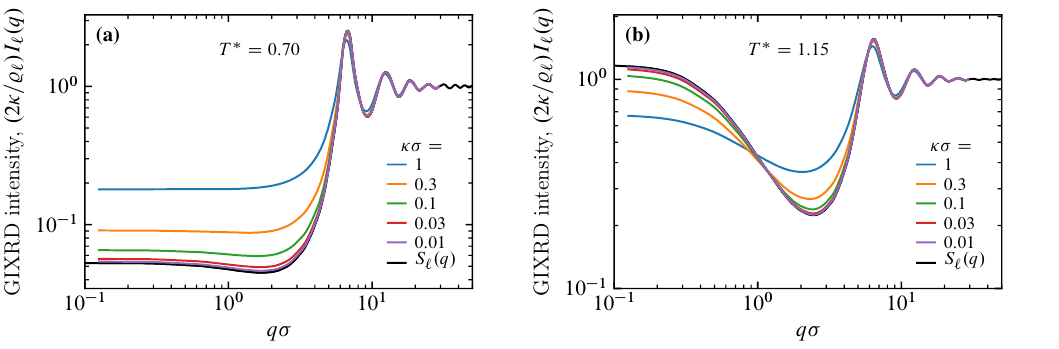}%
  \caption{Contribution of the bulk liquid to the GIXRD intensity for five penetration depths $\kappa^{-1}$ (in units of the LJ diameter $\sigma$) as calculated from \cref{eq:gid_bulk_numerics}.
  As input serve the simulated bulk structure factors $S_\ell(k)$ of the (truncated) LJ liquids at temperatures $T^*=\kB T/\epsilon=0.70$ (a) and $T^*=1.15$ (b) in reduced units with the LJ energy scale $\epsilon$. %, see caption of \cref{fig:ssf_bulk_slab} for details.
  The normalization is taken such that in the limit $\kappa \to 0$ the curves stay finite and approach $S_\ell(q)$ [\cref{eq:gid_bulk_limit}].
  }
  \label{fig:gid_bulk}
\end{figure*}

\begin{figure*}
  \includegraphics[width=1.5\figwidth]{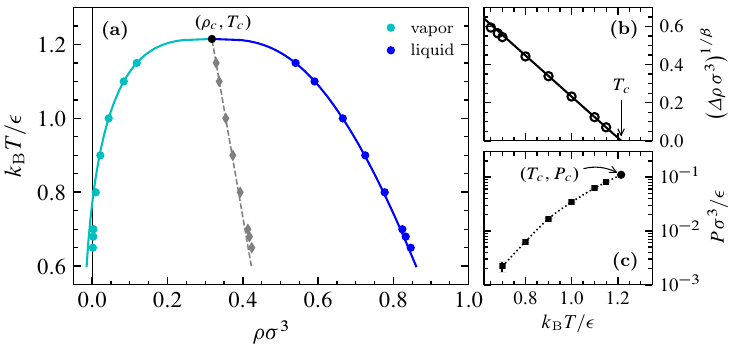}
  \caption{The liquid--vapor transition within the phase diagram of the truncated LJ fluid with interaction cutoff radius $r_c = 3.5\sigma$.
  ~(a)~Binodal line with the coexisting densities $\rho_\ell$ (dark blue disks) and $\rho_v$ (cyan blue disks) taken from fits to the density profile of the inhomogeneous system (\cref{tab:coexistence}).
  Solid lines represent the critical law, $\rho_{\ell/v}(T) \simeq \rho_\mathrm{sym}(T) \pm A_\rho |t|^\beta$ for $t := (T-T_c)/T_c \uparrow 0$ with the universal critical exponent $\beta = 0.325$ of the three-dimensional Ising universality class.
  The gray data points indicate the symmetry line of the binodal, i.e., $\rho_\mathrm{sym} = (\rho_\ell + \rho_v) / 2$;
  its linear extrapolation to $T_c$ (dashed line) yields the critical density $\rho_c$.
  ~(b)~Rectification of the critical law by plotting $(\Delta \rho)^{1/\beta} = (\rho_\ell - \rho_v)^{1/\beta}$ vs. $T$ for the same data as in (a); a linear regression to the three data points for $T^* \geq 1.0$ yields the critical temperature $T_c^*=1.215 \pm 0.001$.
  ~(c)~Coexistence line in the pressure--temperature plane;
  symbols are simulation data with the disk marking the critical pressure $P_c = (0.11074 \pm 0.00003) \epsilon \sigma^{-3}$;
  the dotted line is a smooth interpolation of the simulation data.
  }
%   Δρ(T) = A^β (T - Tc)^β, A = -1.08 ± 0.0049, Tc = 1.215 ± 0.00066
%   Tc - T = B[(ρ_ℓ + ρ_v) / 2 - ρ_c], B = 5.868 ± 0.11, ρc = 0.318 ± 0.00047
  \label{fig:coexistence}
\end{figure*}

The behavior of the bulk contribution $I_\ell(q)$ is presented exemplarily in \cref{fig:gid_bulk} for a Lennard-Jones (LJ) liquid along the liquid--vapor coexistence line at two temperatures: $T^*:=\kB T/\epsilon=0.70$ slightly above the triple point temperature and
$T^*=1.15$ in proximity of the critical temperature ($T_c^* \approx 1.22$).
The pair potential was truncated at $r_c=3.5\sigma$, and $\epsilon$ and $\sigma$ refer to the interaction strength and range of the LJ potential, respectively; the phase diagram is shown in \cref{fig:coexistence} and further details of the simulations are given in \cref{sec:sim_details}.
We calculated $I_\ell(q)$ according to \cref{eq:gid_bulk_numerics} from the bulk structure factors $S_\ell(k)$ as input,
using the same simulation data for $S_\ell(k)$ as the ones in Ref.~\citenum{SSFslab:JCP2020}.
The integration over $k_z$ in \cref{eq:gid_bulk_numerics} was restricted to $0 \leq k_z \leq k_\text{max} = 50/\sigma$, and $S_\ell(k)$ was extrapolated to such large wave numbers as described in Ref.~\citenum{SSFslab:JCP2020}.
The curves obtained for a range of penetration depths ($0.01 \leq \kappa\sigma \leq 1$) convincingly corroborate the convergence of $(2\kappa/\rho_\ell) \, I_\ell(q)$ to $S_\ell(q)$ as $\kappa \to 0$.
The corrections, however, are significant for $\kappa\sigma \gtrsim 0.1$:
from \cref{fig:gid_bulk} one infers that the value of $I_\ell(q\to 0)$ is increased over the value of $S_\ell(q\to 0)$ by a factor of up to $\approx 3.5$ for $T^*=0.70$, which is close to the triple point, whereas it is suppressed by a factor of about 1.8 at the higher temperature ($T^*=1.15$).
Moreover, the minimum in $I_\ell(q)$ near $q\sigma \approx 2$ becomes more shallow for increasing $\kappa$
and seems to disappear at low temperatures.

\subsection{GIXRD intensity from a liquid slab of finite width}
\label{sec:finite_slab}

\begin{figure} \fboxsep=0pt
 \includegraphics[width=.4\textwidth]{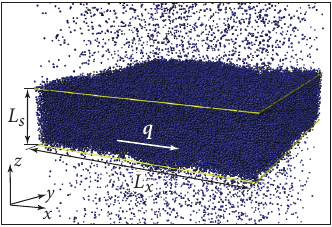}
 \caption{Snapshot of the simulated LJ liquid--vapor coexistence at $T\approx 0.8 T_c$, comprising \num{210000} particles in total; in the vapor phase, only every 10th particle has been drawn for clarity. The yellow frames indicate the mean positions of the two planar interfaces, which delimit the liquid slab of width $L_s = 25.0\sigma$ [see \cref{eq:symmetric_sigmoidal_profile}];
 the area of each of the mean interfaces is $(100\sigma)^2$.
 The lateral scattering vector $\vec q$ lies in the plane parallel to the interfaces ($xy$-plane).
 Along the direction normal to the interfaces ($z$-axis), the length of the simulation box is $L_z=125\sigma$, but only a part of it is shown.
 }
 \label{fig:snapshot-3d}
\end{figure}

For MD simulations of GIXRD intensities, the finite size of the system and, in particular,
the finite width $L$ of the bulk liquid must be accounted for properly (\cref{fig:snapshot-3d}). One anticipates that
the appropriate expression for the bulk scattering combines aspects both of \cref{sec:ssf_slab,sec:half_space}.
We start the derivation of the expression for the bulk scattering by reiterating that the reference system is such that liquid and vapor regions are independent of each other and that there are no distortions of the microscopic density close to $z=-L_\ell, 0, L_v$,
where free boundary conditions are imposed.
The finite widths of liquid and vapor slabs, respectively, are implemented as cutoffs in the weight function:
\begin{equation}
  f(z) = \begin{cases}
    1, & 0 < z \leq L_v \,, \\
    \exp(-\kappa |z|), & -L_\ell \leq z < 0 \,, \\
    0, & \text{otherwise.}
  \end{cases}
  \label{eq:damping_finite}
\end{equation}
Concerning the derivation of a formula for the background scattering, we again specialize to the liquid side ($z < 0$).
The bulk contribution of the vapor side follows in the limit $\kappa \to 0$ as discussed previously \cite{SSFslab:JCP2020} (see \cref{sec:ssf_slab}).

Following the same route as before, we combine the GIXRD master formula [\cref{eq:gid_master}] for the two-point correlation function $G_\ell(q, \Delta z)$ of the homogeneous bulk with the truncated form of $f(z)$ [\cref{eq:damping_finite}] which yields the finite-slab ($L_\ell < \infty$) version of \cref{eq:gid_uv_integrals}:
\begin{align}
  I_\ell(q)
    &= \frac{1}{2}\int\limits_{-L_\ell}^{L_\ell}\!\diff u \!\int\limits_{|u|}^{2L_\ell - |u|} \! \diff v\, \e^{-\kappa v} \, G_\ell(q, u) \notag \\
    &= \frac{1}{\kappa} \int_0^{L_\ell}\!\diff u \,
      2\e^{-\kappa L_\ell} \sinh\boldsymbol(\kappa(L_\ell-u)\boldsymbol) \, G_\ell(q, u) \,,
\end{align}
after carrying out the elementary integral over $v$ in the first line.
In the next step, we use \cref{eq:Gq_bulk} in order to substitute $G_\ell(q, u)$ by the bulk structure factor $S_\ell(k)$ and interchange the integrations over $k_z$ and $u$.
Based on the integral
\begin{multline}
  \frac{1}{\kappa}\int\limits_0^{L_\ell}\!\diff u \, \sinh\boldsymbol(\kappa(L_\ell-u)\boldsymbol)\, \cos(k_z u) \\
  = \frac{\cosh(\kappa L_\ell)-\cos(k_z L_\ell)}{\kappa^2 + k_z^2} \,,
\end{multline}
we arrive at our central result for a finite system:
\begin{align}
I_\ell(|\vec q|) &=
  \rho_\ell \int_{-\infty}^\infty\! \frac{\diff k_z}{2\pi} \,
    \frac{S_\ell\bigl(\sqrt{|\vec q|^2 + k_z^2}\bigr)}{\kappa^2 + k_z^2} \,\times \notag \\
  & \hspace{4em} 2\e^{-\kappa L_\ell} \bigl[\cosh(\kappa L_\ell) - \cos(k_z L_\ell) \bigr] \notag \\
  & \qquad - \frac{\rho_\ell^2}{\kappa^2} \left(1 - \e^{-\kappa L_\ell} \right)^2 (2\pi)^2\delta(\vec q)  \,;
  \label{eq:gid_finite_bulk}
\end{align}
the last line contains the contribution for $\vec q=0$, and its prefactor stems from the integral
\begin{equation}
  \frac{1}{\kappa}\int\limits_0^{L_\ell}\!\diff u \, 2\e^{-\kappa L_\ell} \, \sinh\boldsymbol(\kappa(L_\ell-u)\boldsymbol)
  = \frac{\left(1-\e^{-\kappa L_\ell}\right)^2}{\kappa^2} \,.
\end{equation}
In the final step, we make use of the identity
\begin{equation}
  \int_{-\infty}^\infty\! \frac{\diff k_z}{2\pi} \,
  \frac{\cosh(\kappa L_\ell) - \cos(k_z L_\ell)}{\kappa^2 + k_z^2}  = \frac{\sinh(\kappa L_\ell)}{2\kappa}
  \label{eq:cosh-cos_integral}
\end{equation}
and recast \cref{eq:gid_finite_bulk} in a form similar to \cref{eq:gid_bulk_numerics}:
\begin{align}
I_\ell(q > 0) &=
  \rho_\ell \int_{-\infty}^\infty\! \frac{\diff k_z}{2\pi} \,
    \frac{S_\ell\bigl(\sqrt{q^2 + k_z^2}\bigr) - 1}{\kappa^2 + k_z^2} \,\times \notag \\
  & \hspace{4em} 2\e^{-\kappa L_\ell} \bigl[\cosh(\kappa L_\ell) - \cos(k_z L_\ell) \bigr] \notag \\
  & \qquad + \frac{\rho_\ell}{2\kappa} \left(1-\e^{-2\kappa L_\ell}\right) \,,
  \label{eq:gid_finite_bulk_numerics}
\end{align}
where the 1 in the numerator of the integrand is balanced by the last term.
With this, the remaining integration over $k_z$ is approximated well by a finite integration domain
$|k_z| < k_\text{max}$, because $S(k) \simeq 1$ for $k$ sufficiently large.
Thus, \cref{eq:gid_finite_bulk_numerics} is suitable for a numerical evaluation;
in particular, increasing $k_\text{max}$ decreases the truncation error.
We have followed the procedure described in Ref.~\citenum{SSFslab:JCP2020}, which has already been used to integrate \cref{eq:gid_bulk_numerics}, choosing $k_\text{max}=50/\sigma$ as before.
\footnote{For $k_\text{max}=50/\sigma$, we have checked for the Ashcroft--Lekner model [\cref{eq:ashcroft-lekner}] that the truncation error introduced in the leading \emph{correction} $\mathcal{J}_0(q)$ [\cref{eq:correction_integral}] to $I_\ell(q)$ is less than 5\%.}

It is elucidating to discuss certain limiting cases of
\cref{eq:gid_finite_bulk}.
For large wave number $q$, one reads off from \cref{eq:gid_finite_bulk_numerics}, using that $S_\ell(q\to \infty) = 1$:
\begin{equation}
  I_\ell(q \to \infty)
    = \frac{\rho_\ell}{2\kappa} \left(1 - \e^{-2\kappa L_\ell} \right)
    \to \begin{cases}
      \, \rho_\ell L_\ell \,, & \kappa \to 0 , \\
      \, \rho_\ell / (2\kappa) \,, & L_\ell \to \infty . \\
    \end{cases}
  \label{eq:gid_finite_bulk_large_q}
\end{equation}
In \cref{eq:gid_finite_bulk,eq:gid_finite_bulk_numerics}, care is needed when taking the limits $\kappa\to 0$ and $L_\ell\to \infty$ at finite values of~$q$.
For $\kappa$ fixed, the limit ${L_\ell\to\infty}$ and the integration over $k_z$ may
be interchanged due to the dominated convergence theorem (using that the numerator is bounded), which allows one to reproduce
the previous results in \cref{eq:gid_bulk,eq:gid_bulk_limit}:
\begin{align}
  \lim_{\stackrel{L_\ell \to \infty}{\kappa \:\text{fixed}}} I_\ell(q>0)
  &= \rho_\ell \int_{-\infty}^\infty\! \frac{\diff k_z}{2\pi}
    \frac{S_\ell\bigl(\sqrt{q^2 + k_z^2}\bigr)}{\kappa^2 + k_z^2}  \notag \\
  &\simeq \frac{\rho_\ell}{2\kappa} \, S_\ell(q) \,, \qquad \kappa \to 0.
  \label{eq:gid_finite_bulk_asymptote}
\end{align}

The limit $\kappa\to 0$ for $L_\ell$ fixed is more intricate.
The integrand in \cref{eq:gid_finite_bulk} is dominated by the integrable function
$M k_z^{-2} {[2 - 2 \cos(k_z L_\ell)]}$ for all $\kappa \geq 0$, where $M$ is the maximum of $S_\ell(\cdot)$.
[Proof: Put $y=\e^{-\kappa L_\ell}$ and $a=\cos(k_z L_\ell)$ and use the fact that the expression $1+y^2-2y a$ is monotonically increasing for $y \geq a$, i.e., it is maximal at $y=1$.]
Therefore, the limit $\kappa\to 0$ may be taken inside the integral so that
\begin{multline}
  \lim_{\stackrel{\kappa \to 0}{L_\ell \:\text{fixed}}} I_\ell(q>0) = \\
 2 \rho_\ell L_\ell \int\limits_0^\infty\! \frac{\diff x}{\pi} \,
    \frac{1 - \cos(x)}{x^2} \, S_\ell\bigl(\sqrt{q^2 + (x/L_\ell)^2}\bigr)
  \label{eq:gid_finite_bulk_kappa0}
\end{multline}
where $x = k_z L_\ell$.
Except for the prefactor $\rho_\ell L_\ell$, this is precisely the structure factor of a liquid slab [see \cref{eq:ssf_slab}], and only for large $L_\ell$ it approaches the bulk structure factor:
\begin{align}
  \lim_{\stackrel{\kappa \to 0}{L_\ell \:\text{fixed}}} I_\ell(q>0)
 &= \rho_\ell L_\ell S(q;L_\ell) \notag \\
 &\simeq \rho_\ell L_\ell\, S_\ell(q) \,, \qquad L_\ell \to \infty .
  \label{eq:gid_finite_bulk_kappa0_asymptote}
\end{align}
This result is particularly relevant for simulations, because for them finite-size effects cannot be avoided,
and because a finite value of $L_\ell$ permits to probe the physically meaningful limit $\kappa \to 0$ directly.

\begin{figure*}
  \includegraphics[width=\textwidth]{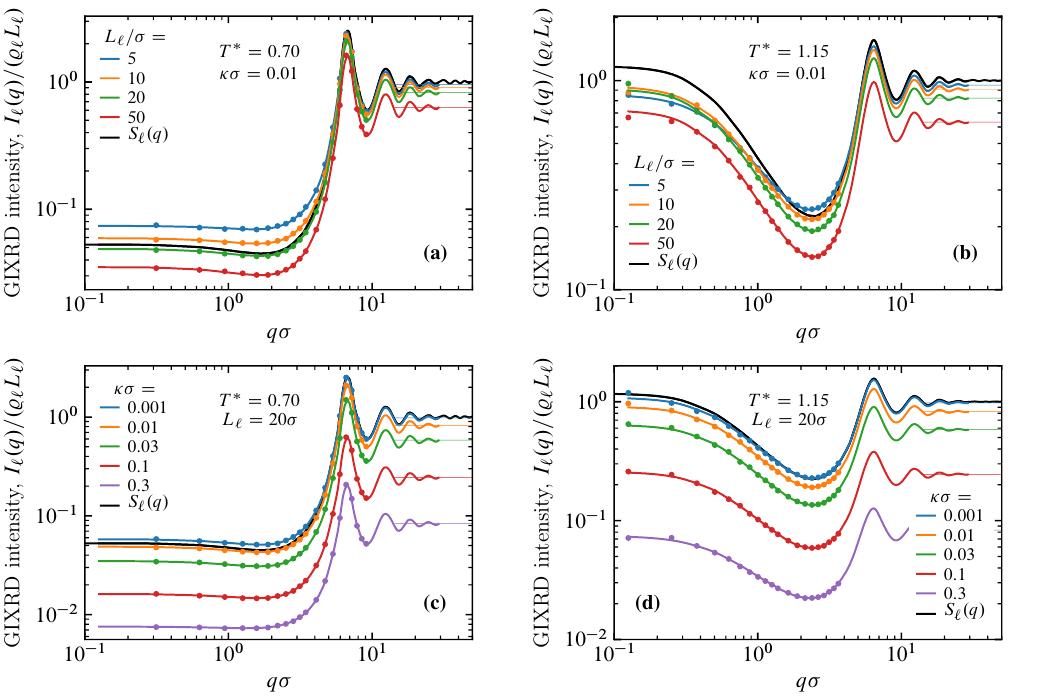}
  \caption{Contribution to the GIXRD intensities collected from liquid slabs of finite width $L_\ell$ and with open boundary conditions.
  Colored solid lines have been obtained from \cref{eq:gid_finite_bulk_numerics} using the bulk structure factor $S_\ell(q)$ (black line) as input; symbols are results which follow from the direct evaluation of \cref{eq:gid_master} within MD simulations for bulk liquids.
  The top panels [(a),(b)] show results for liquid slabs of four widths at fixed inverse penetration depth $\kappa\sigma=0.01$.
  The bottom panels [(c),(d)] show results for a liquid slab of fixed width $L_\ell=20\sigma$ with five penetration depths.
  The left panels [(a),(c)] refer to a liquid at the temperature $T^*=0.70$, i.e., close to the triple point, and
  the right panels [(b),(d)] correspond to $T^*=1.15$ near the liquid--vapor critical point.
  The thin horizontal lines at large $q$ indicate the limits given in \cref{eq:gid_finite_bulk_large_q}.
  }
  \label{fig:gid_finite_bulk}
\end{figure*}

In \cref{fig:gid_finite_bulk}, we test the expression in \cref{eq:gid_finite_bulk} for the GIXRD intensity scattered from finite-sized liquid samples at low ($T^*=0.70$) and high ($T^*=1.15$) temperatures.
As for \cref{fig:gid_bulk}, we have employed the bulk structure factors $S_\ell(q)$ obtained from MD simulations.
Alternatively, we have calculated $I_\ell(q)$ within the simulations and via \cref{eq:gid_master}, with the integration domain restricted to $0 \leq z, z' \leq L_\ell$. These simulation data are in excellent agreement with the results from \cref{eq:gid_finite_bulk} for all parameter sets.
The top row of the panels [(a),(b)] shows the convergence of $I_\ell(q)$ upon increasing the sample width $L \to \infty$ at a fixed, exemplary penetration depth $\kappa \sigma = 0.01$,
whereas in the bottom row [(c),(d)] the limit $\kappa \to 0$ is taken at the fixed sample width $L_\ell=20\sigma$.
We note that in the latter case ($\kappa \to 0, L_\ell \:\text{fixed}$), the scattered intensity does not converge to the bulk structure factor $S_\ell(q)$ (black lines), but rather to the slab structure factor $S(q;L_\ell)$ (not shown), as expected from \cref{eq:gid_finite_bulk_kappa0_asymptote}; the difference is $O(L_\ell^{-1})$ and vanishes for macroscopic samples [see \cref{eq:ssf_slab_asymptotics}].

It is noteworthy that the bulk contribution $I_\ell(q)$ to the scattered intensity, including the case $\kappa=0$, is non-additive:
\begin{equation}
  I_\ell(q;\kappa, L_1) + I_\ell(q;\kappa, L_2) \neq I_\ell(q;\kappa, L_1 + L_2)  \,.
  \label{eq:gid_additivity}
\end{equation}
On the other hand, the expression $\rho_\ell L_\ell\, S_\ell(q)$ is trivially $L_\ell$-additive, implying that
it does not contain correlations in the transversal direction, i.e., between two particles at positions $z\neq z'$.
However, such correlations are contained in the scattered intensity.
One anticipates an appreciable error in the interfacial structure factor $H(q)$ [\cref{eq:H_q}] at small yet non-zero wave numbers $q$ if the bulk contribution $I_\ell(q)$ is approximated by $I_\ell(q) \approx \rho_\ell L_\ell S_\ell(q)$ --- which would correspond to assuming periodic boundary conditions at the interface (see the last paragraph of \cref{sec:ssf_slab}).
A quantification of this error in $H(q)$ follows directly from the expansion of $S(q;L_\ell)$ for large $L_\ell$
[\cref{eq:ssf_slab_asymptotics}], which yields
\begin{equation}
  \lim_{L_\ell \to \infty} \lim_{\kappa \to 0} \left[I_\ell(q) - \rho_\ell L_\ell S_\ell(q) \right]
  = 2 \rho_\ell \mathcal{J}_0(q) \,.
  \label{eq:gid_finite_corr}
\end{equation}
An analogous error in $H(q)$ arises in the analysis of experimental data if only the leading order of the bulk scattering is subtracted from the scattered intensity [see \cref{eq:gid_bulk_corr}].
However, the order of the limits $\kappa \to 0$ and $L_\ell \to \infty$ is different in the two cases, which results in
the prefactor 2 on the r.h.s.\ of \cref{eq:gid_finite_corr} relative to \cref{eq:gid_bulk_corr}. The latter is understood by noting that $\kappa L_\ell$ in \cref{eq:gid_finite_bulk} is sent either to $0$ (here) or to $\infty$ [\cref{eq:gid_bulk_corr}].

\subsection{Uniform asymptotic behavior}

For the interpretation of both experiments and simulations, the asymptotic behavior of the bulk scattering for small (albeit non-zero) $\kappa$ and large (finite or infinite) $L_\ell$ is relevant. Accordingly, the issue arises whether \cref{eq:gid_finite_bulk} can be recast such that the leading asymptotic behavior is apparent from a single expression for both orders of the limits $\kappa\to 0$ and $L\to \infty$.
Inspired by \cref{eq:gid_finite_bulk_large_q} and by the asymptotes of $I_\ell(q)$ [\cref{eq:gid_finite_bulk_kappa0_asymptote,eq:gid_finite_bulk_asymptote}], we single out a term proportional to $S_\ell(q)$ by using the identity in \cref{eq:cosh-cos_integral} and by rearranging the remaining integral:
\begin{align*}
  & \frac{1}{\rho_\ell} I_\ell(q > 0) =
  \frac{1-\e^{-2\kappa L_\ell}}{2\kappa} \, S_\ell(q) \\
  & \qquad + 2\e^{-\kappa L_\ell} \cosh(\kappa L_\ell) \int_0^\infty\! \frac{\diff k_z}{\pi} \,
    \frac{S_\ell\bigl(\sqrt{q^2 + k_z^2}\bigr) - S_\ell(q)}{k_z^2} \,\times \\
  && \mathllap{\left(1-\frac{\kappa^2}{\kappa^2 + k_z^2}\right)\hspace{1em}} \\
  & \qquad - 2\e^{-\kappa L_\ell}  \int_0^\infty\! \frac{\diff k_z}{\pi} \,
    \frac{S_\ell\bigl(\sqrt{q^2 + k_z^2}\bigr) - S_\ell(q)}{\kappa^2 + k_z^2} \, \cos(k_z L_\ell) \,.
\end{align*}
With the definition of $\mathcal{J}_0(q)$ in \cref{eq:correction_integral}, this reduces to
\begin{align}
& \frac{1}{\rho_\ell} \, I_\ell(q>0) =
  \frac{1-\e^{-2\kappa L_\ell}}{2\kappa} \, S_\ell(q)
   + \left(1+\e^{-2\kappa L_\ell}\right) \, \mathcal{J}_0(q) \notag \\
& \quad - \left(1+\e^{-2\kappa L_\ell}\right) \int_0^\infty\! \frac{\diff k_z}{\pi} \,
     \frac{\kappa^2}{\kappa^2 + k_z^2} \, \frac{S_\ell\bigl(\sqrt{q^2 + k_z^2}\bigr) - S_\ell(q)}{k_z^2} \notag \\
& \quad - 2\e^{-\kappa L_\ell} \int_0^\infty\! \frac{\diff k_z}{\pi} \,
    \frac{S_\ell\bigl(\sqrt{q^2 + k_z^2}\bigr) - S_\ell(q)}{\kappa^2 + k_z^2} \, \cos(k_z L_\ell) \,.
  \label{eq:gid_bulk_expansion}
\end{align}
The first term on the r.h.s.\ is the leading order for $\kappa\to 0$ and $L_\ell\to \infty$. One recovers both $I_\ell(q) \simeq S_\ell(q) / (2\kappa)$ and $I_\ell(q) \simeq \rho_\ell L_\ell S_\ell(q)$, depending on the order of the limits. The second term
is of $O(1)$ w.r.t.\ $\kappa\to 0$, $L_\ell\to \infty$;
it contains $\mathcal{J}_0(q)$,
which is independent of $\kappa$ and $L_\ell$ and is given solely by $S_\ell(q)$.

The remaining two terms are higher-order corrections.
For the third term, in the limit $\kappa \to 0$ one finds
\begin{align}
& \int\limits_{-\infty}^\infty\! \frac{\diff k_z}{2\pi} \,
    \frac{\kappa^2}{\kappa^2 + k_z^2} \frac{S_\ell\bigl(\sqrt{q^2 + k_z^2}\bigr) - S_\ell(q)}{k_z^2} \notag \\
& \qquad = \frac{\kappa}{2} \lim_{k_z\to 0} \frac{S_\ell\bigl(\sqrt{q^2 + k_z^2}\bigr) - S_\ell(q)}{k_z^2} + o(\kappa) \notag \\
& \qquad = \frac{\kappa}{4 q} \, S_\ell'(q) + o(\kappa) \,.
\end{align}
Here, we have made use of \cref{eq:Dirac_delta} and of the expansion
\begin{equation}
  S_\ell\Bigl(\sqrt{q^2 + k_z^2}\Bigr) = S_\ell(q) + \frac{k_z^2}{2q} S_\ell'(q) + O\bigl(k_z^4\bigr).
  \label{eq:ssf_small_kz}
\end{equation}
We note that $S_\ell'(q)/q < \infty$ for all $q$, including $q\to 0$.
Thus, the second line of \cref{eq:gid_bulk_expansion} vanishes for $\kappa \to 0$ at $L_\ell$ fixed.

The term in the last line vanishes exponentially if the limit $L_\ell\to \infty$ is taken first at fixed $\kappa$.
For finite $L_\ell$, the remaining integral is bounded for all $\kappa \geq 0$
[e.g., by $\mathcal{J}_0(q)$] and it is the Fourier cosine transform of an integrable and continuous function and thus decays as $L_\ell^{-1}$ for $L_\ell \to \infty$ [see also \cref{eq:ssf_slab_asymptotics}].
The limit $\kappa \to 0$ can be interchanged with the integration over $k_z$. Therefore, the last line of \cref{eq:gid_bulk_expansion} behaves as $\e^{-\kappa L_\ell} O\bigl(L_\ell^{-1}\bigr) [1 + O(\kappa)]$.

In summary, the asymptotic behavior of $I_\ell(q)$ in the joint limit $L_\ell\to \infty$ and $\kappa \to 0$ reads:
\begin{align}
& \frac{1}{\rho_\ell} \, I_\ell(q>0) =
  \frac{1-\e^{-2\kappa L_\ell}}{2\kappa} \, S_\ell(q) \notag \\
& \qquad  + \left(1+\e^{-2\kappa L_\ell}\right) \, \left[\mathcal{J}_0(q) - \frac{\kappa}{4 q} S_\ell'(q) + o(\kappa) \right] \notag \\
& \qquad + \e^{-\kappa L_\ell} O\bigl(L_\ell^{-1}\bigr) [1 + O(\kappa)] \,.
\end{align}
Taking $L_\ell\to \infty$ first, one obtains for large penetration depths (i.e., $\kappa \to 0$)
\begin{equation}
 I_\ell(q>0) =
  \frac{\rho_\ell S_\ell(q)}{2\kappa} + \rho_\ell \mathcal{J}_0(q) - \frac{\kappa \rho_\ell S_\ell'(q)}{4 q} + o(\kappa) \,.
  \label{eq:gid_bulk_kappa_expansion}
\end{equation}
Conversely, for $L_\ell$ large but fixed and for $\kappa \to 0$ one has
\begin{align}
  I_\ell(q>0) = \rho_\ell L_\ell S_\ell(q)  + 2 \rho_\ell \mathcal{J}_0(q) + O\bigl(L_\ell^{-1}\bigr) \,.
  \label{eq:gid_bulk_L_expansion}
\end{align}
The next-to-leading order terms of both expansions depend neither on $\kappa$ nor on $L_\ell$, but differ by a factor of~2.
Approximating the background intensity by the leading order only, as widely done in the analysis of experimental or simulation data (and also in theoretical treatments), leaves a contribution proportional to $\rho_\ell \mathcal{J}_0(q)$, but with a different prefactor, in the expression for the interface correlations. This introduces a systematic inconsistency when comparing results from experiments and simulations. (In the former results, unavoidably one has $\kappa > 0$ and an extrapolation $\kappa \to 0$ is needed; in the latter results, one may put $\kappa = 0$ from the very beginning.)
In particular, we shall find below [\cref{eq:gamma_correction}] that this contribution pollutes the effective surface tension $\gamma(q)$ at the order $q^2$ for small~$q$.

\section{GIXRD from liquid--vapor interfaces}
\label{sec:interfaces}

% \subsection{Composed bulk phases}

For GIXRD scattering from liquid--vapor interfaces, we consider the inhomogeneous reference system as described in \cref{sec:Gb_model}, which is composed of two independent bulk phases at the coexisting densities and which exhibits the absence of correlations across the interface and unperturbed bulk correlations even close to the interface [see \cref{eq:Gb_model}].
The evanescent wave on the liquid side is combined with a propagating
wave on the vapor side so that
\begin{equation}
  f(z) = \begin{cases} 1, & z > 0, \\
    \exp(-\kappa |z|), & z \leq 0,
  \end{cases}
  \label{eq:fz_composed}
\end{equation}
in the master formula for the GIXRD intensity [\cref{eq:gid_master}].
(Here, we do not consider the weak absorption of X-rays.).
Evaluating the master formula with $G(q,z,z')$ and $f(z)$ as in \cref{eq:Gb_model,eq:fz_composed}, respectively, yields the background intensity due to bulk scattering,
\begin{equation}
  I_b(q) = I_\ell(q) + \lim_{\kappa \to 0} I_v(q) \,,
  \label{eq:gid_composed}
\end{equation}
with the liquid and the vapor contributions given by \cref{eq:gid_bulk} and \cref{eq:gid_finite_bulk_kappa0}, respectively.
The vapor contribution becomes particularly relevant at elevated temperatures, i.e., upon approaching the critical point.
We also note that the limit $\kappa\to 0$ of $I_v(q)$ is meaningful only for a finite width $L_v < \infty$ of the vapor phase, as naturally encountered in simulations.
For plotting purposes, we quote the large-$q$ limit, which follows from \cref{eq:gid_composed,eq:gid_finite_bulk_large_q}:
\begin{equation}
  I_b(q\to \infty) = (\rho_\ell/2\kappa) \left(1 - \e^{-2\kappa L_\ell} \right) + \rho_v L_v \,.
  \label{eq:gid_composed_large_q}
\end{equation}

Also for the CW divergence of $G(q,z,z')$ in the classical Wertheim--Weeks theory, the presence of the function $f(z)$ in the integrand of the GIXRD master formula implies a $\kappa$-dependent shift of the scattering amplitude.
This is seen by combining \cref{eq:WertheimWeeks,eq:gid_master} and then inserting \cref{eq:fz_composed}:
\begin{multline}
 I_\text{CWT}(q \to 0) \simeq
  \frac{\kB T }{\gamma_0 \, q^2}
  \left|\int_{-\infty}^\infty \! \diff z \, f(z) \,\rho'(z)\right|^2 \, \\
  = \frac{\kB T}{\gamma_0 \, q^2} \left[
    \rho_v - \rho(0) +
    \int_{-\infty}^0 \! \diff z \,\e^{-\kappa |z|} \, \rho'(z)\right]^2 ,
\end{multline}
where $\rho_v = \rho(z\to\infty)$ and $\rho_\ell = \rho(z \to -\infty)$.
Assuming a sigmoidal interface profile of width $\zeta$,
\begin{equation}
  \rho(z) = \frac{\rho_\ell + \rho_v}{2} -  \frac{\Delta \rho}{2} \, \tanh(z / (2 \zeta)) \/,
  \label{eq:sigmoidal_profile}
\end{equation}
the integral can be expressed in terms of the digamma function $\psi_0(x) = \Gamma'(x)/\Gamma(x)$:
\begin{multline}
  I_\text{CWT}(q \to 0;\kappa) \simeq \frac{\kB T}{\gamma_0 \, q^2} (\Delta \rho)^2 \\ \times \left\{
      1 - \frac{\kappa \zeta}{2}\left[
        \psi_0 \left(\frac{\kappa\zeta}{2} + 1\right) -
        \psi_0 \left(\frac{\kappa\zeta}{2} + \frac{1}{2}\right) \right] \right\}^2 \/.
  \label{eq:I_CWT_corrected}
\end{multline}
Except in the close vicinity of the liquid--vapor critical point, it holds $1/\kappa \gg \zeta$ and one
may approximate the second line by
% $1 - \kappa \zeta [\psi_0(1)-\psi_0(1/2)] + O(\kappa \zeta)^2$
$1 - 1.39 \, \kappa\zeta + O(\kappa \zeta)^2$,
which can be a noticeable correction to \cref{eq:I_CWT} at elevated temperatures;
for the last result, we used $\psi_0(1)-\psi_0(1/2) \approx 1.39$.

\subsection{Wavenumber-dependent surface tension}

For given GIXRD intensities $I(q;\kappa)$ of liquid--vapor interfaces,
the $q$-dependent surface tension follows as the limit $\gamma(q) =  \gamma(q;\kappa \to 0)$, where
\begin{equation}
 \gamma(q;\kappa) := \frac{\kB T(\Delta \rho)^2}{q^2 I_\text{int}(q;\kappa)} \,,
 \label{eq:gid_gamma_q}
\end{equation}
with, after subtracting the background,
$I_\text{int}(q;\kappa)  = I(q;\kappa) - I_b(q;\kappa)$.
The limit $H(q) = I_\text{int}(q;\kappa \to 0)$ remains regular and is referred to as the interface structure factor [see \cref{eq:Iq_sep,eq:H_q,eq:gamma_q_def}].

Let us examine the implications for $\gamma(q)$ upon replacing the background intensity by the leading term, i.e., if only the divergent part of the bulk scattering is subtracted as it has been done in previous experimental studies \cite{Fradin:2000, Mora:2003, Li:2004}.
To this end, we introduce an approximation $\wtilde H$ for the interface structure factor:
\begin{equation}
  \wtilde H(q) = \lim_{\kappa\to 0} \bigl[I(q) - \frac{\rho_\ell}{2\kappa} \,S_\ell(q) \bigr] \,,
  \label{eq:tildeHq}
\end{equation}
and the corresponding surface tension,
\begin{equation}
  \tilde \gamma(q) = \kB T (\Delta\rho)^2 \big/ \bigl[q^2 \wtilde H(q)\bigr] \,.
\end{equation}
\Cref{eq:gid_bulk_asymptotics} implies that $\wtilde H(q) = H(q) + \rho_\ell \mathcal{J}_0(q)$. We emphasize that this latter $q$-dependent shift persists after taking the limit $\kappa \to 0$.
Thus, including the full expression for the background scattering results in a relative change of $\tilde \gamma(q)$ given by
\begin{align}
 \frac{\gamma(q)}{\tilde\gamma(q)} &= \frac{\wtilde H(q)}{H(q)} =
    1 + \frac{q^2 \gamma(q)}{(\Delta\rho)^2 \kB T} \,\rho_\ell \mathcal{J}_0(q) \notag \\
  & \simeq 1 + \frac{\rho_\ell \gamma_0 \mathcal{J}_0(q\to 0)}{(\Delta\rho)^2 \kB T} \,q^2 , \qquad q \to 0.
  \label{eq:gamma_correction}
\end{align}
The magnitude of this shift of the contribution $O(q^2)$ to $\gamma(q)$ is determined by two lengths:
$\rho_\ell \gamma_0 /[(\Delta\rho)^2 \kB T]$ and $\mathcal{J}_0(q\to 0)$, both of which are amenable to experimental investigations.
As an example, we have estimated these quantities for liquid water at $T=\SI{27}{\degreeCelsius}$, where $\Delta\rho \approx \rho_\ell$, and have obtained
$\gamma_0 /(\rho_\ell \kB T) \approx \SI{5.2}{\angstrom}$
and $\mathcal{J}_0(q \to 0) \approx \SI{0.22}{\angstrom}$
from evaluating \cref{eq:correction_integral} for available data of the bulk structure factor \cite{Clark:MP2010, Hura:PCCP2003}.

In the following, we shall elucidate this shift further for simple theoretical models, which support the simulation results for LJ fluids reported in Ref.~\citenum{Capillary:2015} and below in \cref{sec:MD}.

\subsection{Simple DFT models: square-gradient approximation}

\begin{figure}
 \includegraphics[width=\figwidth]{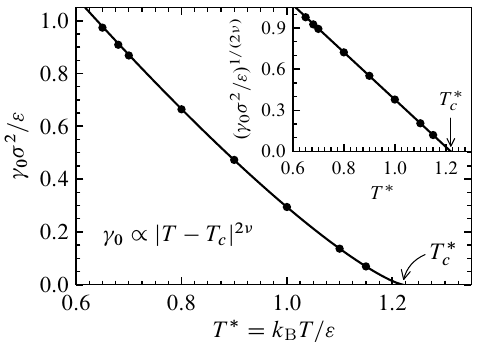}
 \caption{Temperature dependence of the macroscopic surface tension $\gamma_0$ along the liquid--vapor coexistence line from the triple point temperature $T_t^* \approx 0.65\dots 0.70$ to the critical temperature $T_c^*$.
 The data points stem from MD simulations for truncated LJ fluids ($r_c = 3.5\sigma$, \cref{tab:interface}).
 The solid line depicts the critical scaling law $\gamma_0 \simeq A_\gamma |t|^{2\nu}$ for $t:=(T-T_c)/T_c$ upon $t \uparrow 0$ with the Ising exponent $\nu = 0.630$.
 The inset shows the same data in a rectification plot of the critical law, yielding the critical temperature $T_c^* = 1.220 \pm 0.001$ and the amplitude $A_\gamma = (2.554 \pm 0.008) \epsilon \sigma^{-2}$ from a linear regression to the three data points for $T^*\geq 1.0$.
 }
% γ(T) = A (T - T0), A = -1.727 ± 0.0018, T0 = 1.219 ± 0.00021  if T ≥ 0.9
% γ(T) = A (T - T0), A = -1.725 ± 0.0038, T0 = 1.220 ± 0.00033  if T ≥ 1.0
% A_γ = (A × T_c)^2ν = 2.554, the combined relative error is less than 0.003
 \label{fig:gamma_crit}
\end{figure}

The Ornstein--Zernike form
\begin{equation}
 S_\text{OZ}(k) = \frac{S_0}{1+(k\xi)^2}
 \label{eq:ornstein-zernike}
\end{equation}
of the structure factor is a common feature of density-functional theories based on the square-gradient approximation;
here $S_0 = \rho_\ell \kB T \chi_T$, in terms of the isothermal compressibility $\chi_T$, and $\xi$ is the OZ correlation length.
$S_\text{OZ}(k)$ is a useful approximation for the structure factor of liquids within the range $k\xi \ll 1$ and at elevated temperatures. (This is valid for liquids with an appreciable compressibility, but not too close to their critical point.)
For the integral corresponding to the leading-order correction [\cref{eq:correction_integral}] one finds \cite{SSFslab:JCP2020}
\begin{equation}
  \mathcal{J}_0(q) = -\frac{(\xi/2) S_0}{\bigl[1 + (q\xi)^2\bigr]^{3/2}}\,.
  \label{eq:correction_integral_OZ}
\end{equation}
Inserting this into \cref{eq:gid_bulk_asymptotics} yields, for the small-angle bulk scattering,
\begin{equation}
  I_\ell(q\to 0) \simeq \frac{\rho_\ell}{2\kappa} \,S_0  - (\rho_\ell)^2 \kB T \chi_T \xi / 2.
\end{equation}
This decreases the uncorrected, $q$-dependent surface tension $\tilde \gamma(q)$ [\cref{eq:gamma_correction}]:
\begin{equation}
  \gamma(q \to 0) \simeq \tilde\gamma(q) \left[1 - \left(\frac{\rho_\ell}{\Delta\rho}\right)^2 \frac{\gamma_0 \xi \chi_T}{2} \,q^2 \right] \,.
  \label{eq:gamma_corr_OZ}
\end{equation}
The prefactor of the contribution $O(q^2)$ has the dimension of a length squared and, upon approaching the critical point [$t := (T-T_c)/T_c \uparrow 0$], it diverges as
\begin{align}
  (\Delta\rho)^{-2} \gamma_0 \xi \chi_T
  &\sim |t|^{-2\beta} |t|^{(d-1)\nu} |t|^{-\nu} |t|^{-\gamma} \notag \\
  &\sim |t|^{-2\nu}
\end{align}
for $d$ bulk dimensions and by using the exponent relation \cite{Pelissetto:2002} $2\beta + \gamma = \nu d$.
Thus, the correction to $\tilde \gamma(q)$ scales as $(q\xi)^2$ in square-gradient models.
Noting that the macroscopic surface tension $\gamma_0 = \gamma(q\to 0)$ vanishes as \cite{Jasnow:1984} $\gamma_0 \sim |t|^{2\nu}$ (see \cref{fig:gamma_crit,tab:interface}), one finds that for $T\uparrow T_c$ at $O(q^2)$ the difference between $\gamma(q)$ and $\tilde\gamma(q)$ becomes independent of temperature.

\begin{table}
% define centred columns of flexilbe width
\renewcommand\tabularxcolumn[1]{>{\hfill}p{#1}<{\hfill\hbox{}}}
\heavyrulewidth=.1em
\tabcolsep=.5ex
\small
\begin{tabularx}{\figwidth}{XXXX}
\toprule
% use X-columns to adjust column widths to overall text width,
% centre first part of the table rather than to align on the decimal point
$\kB T/\epsilon$ &
$\gamma_0/\bigl(\epsilon \sigma^{-2}\bigr)$ &
$\ell / \sigma$ &
$\zeta / \sigma$ \\
\midrule[\heavyrulewidth]
% temp  gamma  ell  zeta
0.70 &  0.868(2) & 0.23(2) & 0.528(2) \\
0.80 &  0.664(2) & 0.26(2) & 0.642(2) \\
0.90 &  0.473(1) & 0.37(2) & 0.802(2) \\
1.00 &  0.294(2) & 0.53(2) & 1.048(3) \\
1.10 &  0.137(2) & 0.85(2) & 1.548(4) \\
1.15 &  0.069(2) & 1.24(2) & 2.123(5) \\
\bottomrule
\end{tabularx}
\caption{Interfacial properties of truncated LJ fluids ($r_c=3.5\sigma$) along the liquid--vapor coexistence line.
The surface tension $\gamma_0$ was calculated from the anisotropy of the microscopic stress tensor.
The length $\ell$ quantifies the contribution $O(q^2)$ to $\gamma(q)$ and was obtained from fits of \cref{eq:gamma_q_ell} to the data in \cref{fig:gamma_q}(a).
The interfacial width $\zeta$ was determined from fits of \cref{eq:symmetric_sigmoidal_profile} to the density profile of inhomogeneous systems with a mean interfacial area of $(100\sigma)^2$.
The numbers in parentheses give the statistical uncertainty in the last digit.
}
\label{tab:interface}
\end{table}

The relevance of the correction in \cref{eq:gamma_correction} can be assessed by the comparison with the small-$q$ behavior of $\gamma(q)$. To this end, we consider a simplified DFT treatment of the liquid--vapor interface based on the square-gradient approximation with the symmetric double-parabola potential \cite{Parry:2014}.
By defining $\gamma(q)$ via the scattered intensity, for $\kappa\to 0$ \citet{Parry:2014} obtained
$\tilde\gamma(q \to 0) / \gamma_0 \simeq 1 + \frac{5}{4}(q\xi)^2$.
Accounting for the correction of the bulk scattering due to the open boundary [\cref{eq:gamma_corr_OZ}] subtracts the contribution $\frac{1}{4}(q\xi)^2$, i.e., it decreases the contribution $O(q^2)$ by 20\%, leading to
$\gamma(q \to 0) / \gamma_0 \simeq 1 + (q\xi)^2$.
We note that adopting a different definition of the $q$-dependent surface tension which is solely based on the two-point density correlation function $G(q,z,z')$ would yield a different prefactor at order $O(q^2)$, and such a definition would not require as a prerequisite a model for the bulk scattering as input \cite{Parry:2014}. However, up to date, experimentally $G(q,z,z')$ is not directly accessible.

\subsection{Hard-sphere approximation for the bulk liquid}
\label{sec:AL-model}

At temperatures close to the triple point, the liquid phase is characterized by a low compressibility and a small correlation length; in particular, the bulk structure is dominated by the interparticle repulsion.
In order to estimate the value of $\mathcal{J}_0(q\to 0)$ in this regime, we approximate the liquid bulk structure factor by the one of the Ashcroft--Lekner (AL) model for hard spheres \cite{Ashcroft:1966};
% \cite{Ailawadi:1973} improves the model using the Carnahan-Starling equation of state, however, this introduces an unknown parameter, and \cite{Evans:1981} say that the modification does not lead to a substantial improvement
for simplicity, we ignore the contribution due to the attractive part of the pair potential \cite{Evans:1981}.
We recall that hard spheres do not form liquid or vapor phases; accordingly there is no liquid--vapor interface. Nonetheless, one can expect that the AL model for $S_\ell(k)$ renders a useful estimate of the bulk correction $\mathcal{J}_0(q)$ for dense and nearly incompressible liquids.

\begin{figure}
 \includegraphics[width=\figwidth]{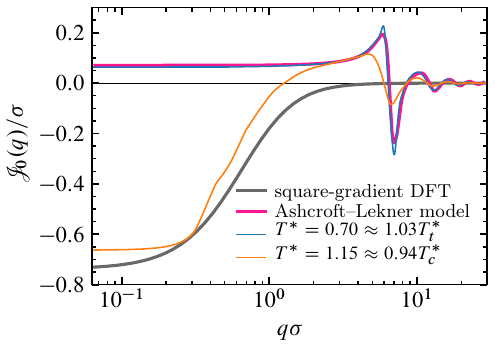}
 \caption{Leading-order correction $\mathcal{J}_0(q)$ to the bulk background [\cref{eq:gid_bulk_asymptotics}] as obtained from quadratures of \cref{eq:correction_integral} with bulk structure factors $S_\ell(k)$ taken from
 (i) the square-gradient DFT [\cref{eq:correction_integral_OZ}],
 (ii) the Ashcroft--Lekner model (\cref{sec:AL-model}),
 and MD simulations of a LJ liquid at the temperatures (iii) $T^*=1.15$ (close to the critical one, $T_c^*$)
 and (iv) $T^*=0.70$ (close to the triple point $T_t^*$).
 The parameters of the theoretical models were chosen to correspond to the simulated liquids:
 (i) $S_0=1.18$ and $\xi=1.25\sigma$ for the DFT model and (ii) $\bar\sigma = \sigma$ at packing fraction $\eta=0.45$ for the hard sphere model.
 Parts of the figure are reproduced from Ref.~\citenum{SSFslab:JCP2020}.
 }
 \label{fig:correction_integral}
\end{figure}

In terms of the volume packing fraction $\eta$ and the hard sphere diameter $\bar\sigma$,
% $\eta = (\pi/6)n\bar\sigma^3$
the AL model reads $S_\text{AL}(k) = [1 - \eta c_\text{AL}(k)]^{-1}$ with the direct correlation function
\begin{gather}
 c_\text{AL}(k) = -24 \int_0^1 \frac{\sin(s k \bar\sigma)}{s k \bar\sigma}(a_0 + a_1 s + a_3 s^3) \, s^2 \diff s
 \label{eq:ashcroft-lekner}
\intertext{and the coefficients}
a_0 = \frac{(1 + 2 \eta)^2}{(1 - \eta)^4}, \:
a_1 = -6 \eta \frac{(1 + \eta / 2)^2}{(1 - \eta)^4}, \:
a_3 = \frac{\eta}{2} \frac{(1 + 2 \eta)^2}{(1 - \eta)^4}.
\end{gather}
% It yields $S_\text{AL}(k\to 0) = (\eta-1)^4/(2 \eta+1)^2$ and, numerically,
Numerical integration of \cref{eq:correction_integral} for $q=0$ yields
the values of $\mathcal{J}_0(0)$, which vary smoothly as function of the packing fraction, attaining their maximum $\approx 0.087\bar\sigma$ near $\eta \approx 0.2$; at high packing fraction $\mathcal{J}_0(0) \approx 0.070 \bar\sigma$ for $\eta=0.45$.
The latter result should be compared with $\mathcal{J}_0(0) \approx 0.063 \sigma$ obtained from MD simulations for a LJ liquid at $T^*=0.70$ (see below for details).
Moreover, using $S_\text{AL}(k)$ as input to \cref{eq:correction_integral} and numerically computing the full $q$-dependence of $\mathcal{J}_0(q)$ yields a remarkably accurate approximation of $\mathcal{J}_0(q)$ with $S_\ell(k)$ obtained from the simulations (\cref{fig:correction_integral}); for this comparison we used $\eta=0.45$ and identified $\bar\sigma$ with $\sigma$.
In particular, our analysis for the OZ and the AL model has revealed that $\mathcal{J}_0(0)$ changes sign due to a subtle competition of excluded volume and long-ranged correlations.

\subsection{Molecular dynamics simulations of Lennard-Jones fluids}
\label{sec:MD}

Within MD simulations, we have numerically determined GIXRD intensities due to scattering off the liquid--vapor interface, based on \cref{eq:gid_master} and a microscopic expression for $G(q,z,z')$; details are given in \cref{sec:sim_details}.
For the calculation of $\gamma(q)$, one has to account for the finite width of the liquid and vapor regions; close to $T_c$, also the vapor contribution to the bulk scattering must not be neglected.
On the other hand, the finite sizes of the bulk phases ensure that both $I(q;\kappa)$ and $I_b(q;\kappa)$ remain finite as $\kappa \to 0$ so that one can put $\kappa=0$ already when calculating $I(q;\kappa)$.
In this case, the dimensionless quantity
\begin{equation}
  S_\text{tot}(q) := \frac{A}{N} \, I(q;\kappa=0)
  \label{eq:Stot-def}
\end{equation}
is given by the standard microscopic expression for the static structure factor [cf.~\cref{eq:I_microscopic} for $f(z)=1$], and the bulk contribution follows from the simpler expression in \cref{eq:ssf_slab}, which was derived in Ref.~\citenum{SSFslab:JCP2020}.
In \cref{eq:Stot-def}, $A$ is the area of the planar mean interface and $N$ is the number of particles in the simulation.
We have followed both routes in order to test their consistency: (i) determine $I(q;\kappa)$ and thus $\gamma(q;\kappa)$ for a decreasing sequence of values of $\kappa$ and take $\gamma(q) = \gamma(q;\kappa\to 0)$, and (ii) calculate $S_\text{tot}(q)$ and thus $\gamma(q)$ directly.

\begin{figure*}
  \includegraphics[width=\linewidth]{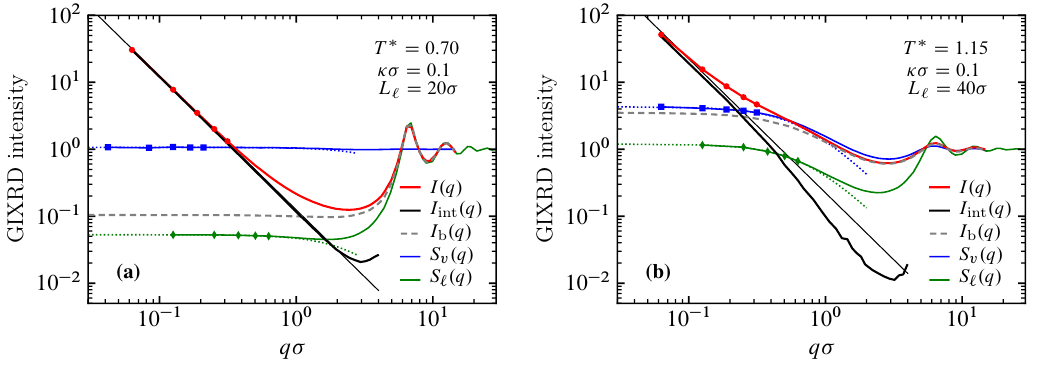}
  \caption{Simulated GIXRD intensity $I(q;\kappa) = I_\text{int}(q;\kappa) + I_b(q;\kappa)$ (red line) from scattering off the corresponding liquid--vapor interface and its decomposition into
  bulk and interface contributions $I_\textrm{b}(q;\kappa)$ and $I_\text{int}(q;\kappa)$, respectively (gray dashed and thick black lines);
  all quantities shown are normalized by $I_b(q \to \infty)$ [\cref{eq:gid_composed_large_q}].
  The two panels show results at the reduced temperatures (a)~$T^*=0.70$ and (b)~$T^*=1.15$, both for the penetration depth $1/\kappa=10\sigma$.
  The thin black line indicates the CW divergence, $I(q \to 0;\kappa) \sim 1/(\gamma_0 q^2)$ [\cref{eq:I_CWT_corrected}], with the macroscopic surface tension $\gamma_0$ obtained independently from the simulated stress tensor [\cref{fig:gamma_crit,tab:interface}];
  the deviation of $I_\text{int}(q;\kappa)$ from the CW divergence (thick vs. thin black lines), which is well developed in panel (b), gives rise to the wave number-dependent surface tension $\gamma(q)$.
  The bulk contribution $I_\textrm{b}(q;\kappa)$ (gray dashed line) was calculated according to \cref{eq:gid_composed,eq:gid_finite_bulk_numerics},
  using the simulated bulk structure factors of the coexisting liquid (green line) and
  vapor (blue) phases $S_\ell(q)$ and $S_v(q)$, respectively; the data are extrapolated to small $q$ assuming the Ornstein--Zernike form (green and blue dotted lines).
  For the calculation of $I(q;\kappa)$ only particles within a slab of width (a) $L_\ell=20\sigma$ and (b) $L_\ell=40\sigma$, respectively, were considered on the liquid side.
  For the quantities $I(q)$, $S_\ell(q)$, and $S_v(q)$, lines connect actual simulation data (symbols, only shown for the five smallest wave numbers).
  In panel (b), the tiny wiggles in $I_\text{int}(q)$ at $q\sigma \gtrsim 1$ reflect the statistical uncertainty of the simulation data.
  }
  \label{fig:gid_analysis}
\end{figure*}

In \cref{fig:gid_analysis}, the decomposition of the scattered intensity into interfacial and background contributions [\cref{eq:Iq_sep}] is illustrated (a) for the temperature $T^*=0.70$ close to the triple point and
(b) for the temperature $T^*=1.15$ in proximity of the liquid--vapor critical point ($T^*_c\approx 1.22$),
both for a penetration depth $1/\kappa = 10\sigma$.
(The analogous decomposition of $S_\text{tot}(q)$ for $T^*=1.15$ is provided by Fig.~2 in Ref.~\citenum{Capillary:2015}.)
For $T^*=0.70$, a liquid slab of width $L_s = 25\sigma$ was simulated, but only particles with positions $0 \leq z \leq L_\ell=20\sigma$ were admitted for the calculation of $I(q;\kappa)$; here, $z=0$ denotes the mean position of the interface.
For the higher temperature, these values have been doubled in order to accommodate the much lower (macroscopic) surface tension and thus larger fluctuations of the local interface position.
On the vapor side, only particles within slabs of $L_v=50\sigma$ ($T^*=0.70$) and $L_v = 75\sigma$ ($T^*=1.15$) were taken into account.
The background scattering was calculated according to \cref{eq:gid_finite_bulk_numerics} using the bulk structure factors $S_\ell(q)$ and $S_v(q)$ as input, which were obtained from separate simulations of homogeneous fluids (see \cref{fig:gid_finite_bulk}).
Close to the triple point [\cref{fig:gid_analysis}(a)], the CW divergence is clearly visible in the scattered intensity, $I(q \to 0;\kappa) \sim q^{-2}$, without adjusting any parameters.
To this end, the prefactor of the divergence was taken from \cref{eq:I_CWT_corrected} using the macroscopic surface tension $\gamma_0$ (\cref{fig:gamma_crit,tab:interface}) as determined independently from an integral over the stress tensor profile across the interface~\cite{Evans:1979,Dunikov:2001}; the interfacial width $\zeta$ was obtained from the simulated mean density profiles $\rho(z)$ [see \cref{eq:symmetric_sigmoidal_profile}].
For the interfacial part of the scattering $I_\text{int}(q;\kappa) \approx H(q)$, this behavior of the CW divergence extends to a wide range of wave numbers $q\sigma \lesssim 2$.
At high temperature [\cref{fig:gid_analysis}(b)], the CW divergence is barely visible in $I(q;\kappa)$ itself, but it can clearly be recognized in $I_\text{int}(q;\kappa)$ for $q\sigma \lesssim 0.2$.
The slight mismatch between the predicted and the actual prefactors of the CW divergence (compare the thin and the thick black lines for $q\sigma  \lesssim 0.2$) disappears for larger penetration of the liquid side, e.g., $\kappa\sigma = 0.01$.
The mismatch is presumably due to higher order terms in \cref{eq:WertheimWeeks};
the apparently obvious cause, that large-amplitude CWs are not probed properly for insufficiently small $\kappa$, is already accounted for in \cref{eq:I_CWT_corrected}.
The sizable deviations of $I_\text{int}(q;\kappa)$ from the asymptotic behavior at large wave numbers give rise to the $q$-dependent surface tension.

\begin{figure}
  \includegraphics[width=\figwidth]{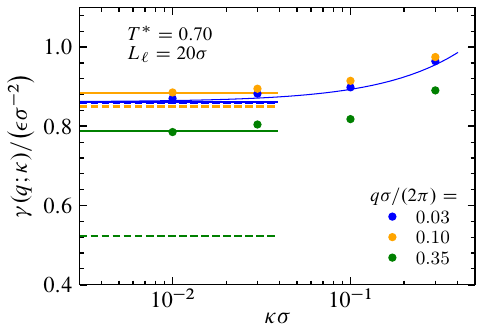}
  \caption{Convergence of the effective $q$-dependent surface tension $\gamma(q;\kappa \to 0)$ as function of the inverse penetration depth $\kappa$ for three, fixed wave numbers $q$ at the temperature $T^*=0.70$.
  The data for $\gamma(q;\kappa)$ were obtained via \cref{eq:gid_gamma_q} from MD simulations for the GIXRD intensity, using only a a fraction $L_\ell=20\sigma$ of the width $L_s$ of the bulk liquid in order to calculate $I(q)$ and carrying out the decomposition shown in \cref{fig:gid_analysis}(a).
  Thick solid lines indicate the corresponding limits $\gamma(q)$ calculated from $S_\text{tot}(q)$, with setting $\kappa=0$ directly within the simulations.
  Dashed lines show $\hat\gamma(q)$ as obtained from $S_\text{tot}(q)$ when accounting only for the divergent part of the bulk scattering [see the main text and \cref{eq:hatHq}].
  The thin blue line ($q\sigma/2\pi = 0.1$) represents the $\kappa$-dependence of $\gamma(q\to 0; \kappa)$ as implied by \cref{eq:I_CWT_corrected}.
  }
  \label{fig:gid_gamma_q_convergence}
\end{figure}

\Cref{fig:gid_gamma_q_convergence} exhibits, for each wave number $q$, the convergence of $\gamma(q;\kappa)$ to the physically meaningful limit $\gamma(q)$ upon systematically decreasing the inverse penetration depth $\kappa \to 0$.
In simulations, $\gamma(q)$ can be obtained also directly from $S_\text{tot}(q)$ (thick solid lines).
However, if only the leading order of the background scattering is subtracted,
\begin{equation}
  \what H(q) := (N/A) S_\text{tot}(q) - [\rho_\ell L_\ell S_\ell(q)+\rho_v L_v S_v(q)] \,,
  \label{eq:hatHq}
\end{equation}
the resulting $\hat\gamma(q)$ deviates from $\gamma(q) = \gamma(q;\kappa \to 0)$ as one would infer from the GIXRD data.
In particular, from \cref{eq:ssf_slab_asymptotics} one concludes that $\what H(q) = H(q) + 2 \rho_\ell \mathcal{J}_0(q)$,
which also differs from $\wtilde H(q)$ [see \cref{eq:tildeHq} and the following text].
Thus,
the discrepancy between $\hat\gamma(q)$ and $\gamma(q)$ is larger for larger wave numbers [cf.~\cref{eq:gamma_correction}];
in the simulations, e.g., for $T^*=0.70$ and $q\sigma \approx 2.2$, we find that $\hat\gamma(q)$ is almost 40\% smaller than $\gamma(q)$.

It was this inconsistency between the analysis of scattering data (as obtained from experiments) and the total structure factor (as obtained within MD simulations) which gave rise to the refined treatment of the bulk background as elaborated here.
The inconsistency would not be lifted by
considering only the divergent part of the background in the analysis of GIXRD intensities $I(q;\kappa)$,
which upon $\kappa \to 0$ would yield $\tilde\gamma(q)$ as introduced after \cref{eq:tildeHq}.
In retrospective, the issue is well understood by comparing
\cref{eq:gid_bulk_kappa_expansion,eq:gid_bulk_L_expansion}, which imply that
$\hat\gamma(q) \neq \tilde\gamma(q)$. This clarifies that the $O(1)$-term $\propto \rho_\ell \mathcal{J}_0(q)$ must be included in the background contribution for a consistent analysis of GIXRD data.
This yields $\gamma(q)$ irrespective of whether it was calculated along route (i) or route (ii), described at the end of the first paragraph in \cref{sec:MD}.

\begin{figure*}
  \includegraphics[width=.5\linewidth]{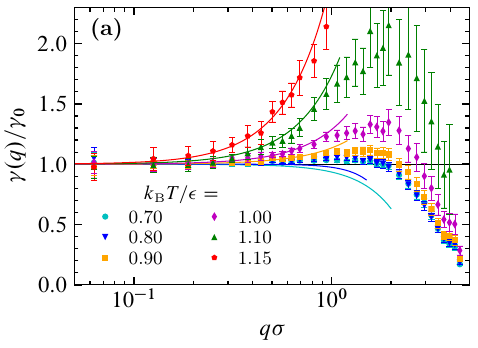}%
  \includegraphics[width=.5\linewidth]{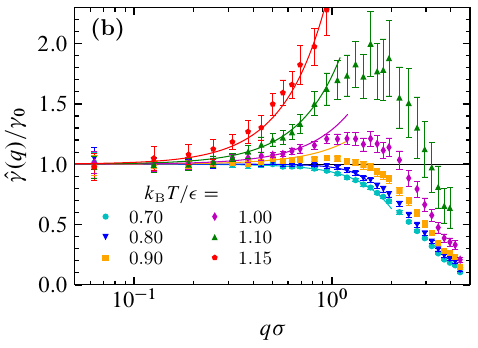}
  \caption{MD simulation results (symbols) for the wave number-dependent surface tension $\gamma(q)$ of the truncated LJ liquid as obtained from $S_\text{tot}(q)$ after subtracting (a) the full background contribution [\cref{eq:ssf_composed}] and (b) only its divergent part [\cref{eq:hatHq}];
  the data points in panel (a) are reproduced from Ref.~\citenum{Capillary:2015}.
  The differences are most apparent at the two lowest temperatures.
  In order to facilitate the comparison, the solid lines are the same in both panels: for $T^* \geq 0.9$, the lines show fits of the small-$q$ behavior as in \cref{eq:gamma_q_ell} to the data for $\gamma(q)$ (panel (a)); for $T^*=0.7$ and $0.8$, they represent empirical power law fits to the data for $\hat\gamma(q)$ (panel (b)).
  % T=0.80: exponent 4
  % T=0.70: exponent 2.5
  }
  \label{fig:gamma_q}
\end{figure*}

The $q$-dependent surface tension $\gamma(q)$ of LJ fluids is shown in \cref{fig:gamma_q}(a) for temperatures ranging from $T^*=0.70$ to $1.15$. The data are taken from Ref.~\citenum{Capillary:2015} and were obtained from the MD simulation results for $S_\text{tot}(q)$ [route (ii)] and from the full background contribution given in \cref{eq:gid_finite_bulk_kappa0_asymptote} and based on \cref{eq:ssf_slab}:
\begin{equation}
  (N/A) S_b(q) = \rho_\ell L_\ell S_\ell(q; L_\ell) + \rho_v L_v S_v(q; L_v) \,.
  \label{eq:ssf_composed}
\end{equation}
The most notable effect is the enhancement of the effective surface tension at non-zero wave numbers upon increasing the temperature; this was discussed in Ref.~\citenum{Capillary:2015}. It can be rationalized by writing
\begin{equation}
  \gamma(q \to 0;T) \simeq \gamma_0(T) \left[1 + q^2\ell(T)^2 \right]
  \label{eq:gamma_q_ell}
\end{equation}
with a temperature-dependent length $\ell(T)$, which increases monotonically as $T$ is increased (\cref{tab:interface}).
% from $(0.23\pm0.02) \sigma$ at $T^*=0.70$ to $(1.24 \pm 0.02) \sigma$ at $T^*=1.15$.
This form of $\gamma(q)$ is corroborated within a recent DFT treatment of liquid--vapor interfaces \cite{Parry:2016,Parry:NP2019,Parry:PRE2019}, with $\ell(T) \simeq \xi(T)$ at temperatures close to the critical one.

\begin{figure}
 \includegraphics[width=\figwidth]{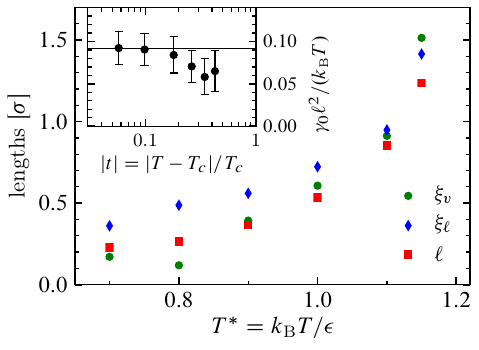}
 \caption{Temperature dependence of the length $\ell$ governing the small-wave number behavior of $\gamma(q\to 0) \simeq \gamma_0[1 + (q\ell)^2]$ in comparison to the bulk correlation lengths, $\xi_\ell$ and $\xi_v$, of the coexisting liquid and vapor phases (\cref{tab:interface}).
 The inset tests the convergence $\gamma_0 \ell^2 / (\kB T) \to (4\pi \omega)^{-1} \approx 0.09$ as $T \to T_c$, where $\omega \approx 0.87$ is a universal amplitude ratio \cite{Mon:PRA1985,Hasenbusch:1993,Das:2011}.
 }
 \label{fig:ell_xi}
\end{figure}

A related observation was made for the curvature-dependence of the macroscopic surface tension, upon replacing $1/q$ by the radius of a spherical droplet \cite{Das:2011}.
The simulation results for $\ell(T)$ exhibit a temperature dependence similar to the one of the OZ bulk correlation lengths of the
coexisting liquid and vapor phases (\cref{fig:ell_xi}).
Moreover, anticipating the same critical scaling exponent as for the correlation length, $\ell(T \uparrow T_c) \sim |T-T_c|^{-\nu}$, the product $\gamma_0 \ell^2$ is expected to converge to a constant.
Interestingly, our data suggest that $\gamma_0 \ell^2 / (\kB T) \to (4\pi \omega)^{-1} \approx 0.09$,
where $\omega = \lim_{T\uparrow T_c} \kB T_c / (4\pi \gamma_0 \xi_{\ell,v}^2)$ is a universal amplitude ratio \cite{Mon:PRA1985,Hasenbusch:1993,Das:2011}; its most reliable estimate stems from Monte Carlo simulations of the three-dimensional Ising model \cite{Hasenbusch:1993}: $\omega \approx 0.87$.
This would imply that indeed $\ell(T) / \xi_{\ell,v}(T) \to 1$ as $T \uparrow T_c$.

The right panel of \cref{fig:gamma_q}(b) shows the corresponding results for $\hat\gamma(q)$, which have been obtained from the same input data, but taking into account only the divergent part of the bulk scattering [\cref{eq:hatHq}].
(We recall that $\mathcal{J}_0(0)$ changes sign as function of temperature, which is seen in \cref{fig:correction_integral}).
Whereas the data points shift slightly upwards at higher temperatures, as expected from \cref{eq:gamma_correction} due to $\mathcal{J}_0(0) < 0$ at these temperatures [\cref{eq:correction_integral_OZ}], the repercussions are more significant at low temperatures ($T^* \lesssim 0.80$):
opposed to the almost constant behavior of $\gamma(q)$ up to $q\sigma \lesssim 2$,
$\hat\gamma(q)$ bends downwards, which results from
$\mathcal{J}_0(0) \approx 0.063 \sigma > 0$ in this case and due to the small value of $\ell$.
Empirically, the data are described by $\hat\gamma(q) \approx \gamma_0(1 + \mathcal{K} q^\alpha)$ with $\mathcal{K} < 0$ and exponents $\alpha = 4$ for $T^*=0.80$ and $\alpha = 2.5$ for $T^*=0.70$.
In particular, the data for $\hat\gamma(q)$ suggest that there is a distinguished temperature $T_0$ with $0.80 \lesssim T_0^* \lesssim 0.90$ such that
$\ell(T) = 0$ for $T < T_0$,
which appears to be implausible on physical grounds.
This issue is removed by considering the full background scattering [\cref{eq:ssf_composed}], which includes the correction given by $\mathcal{J}_0(q)$ and which leads to $\gamma(q)$ as shown in \cref{fig:gamma_q}(a).

\subsection{Sensitivity to the mean interface position}
\label{sec:GDS}

So far, for the interpretation of the simulation data, we have anticipated that the widths $L_\ell$ and $L_v$ of the coexisting phases [\cref{eq:damping_finite}] are known.
Our protocol to construct the equilibrated inhomogeneous samples (see \cref{sec:sim_details}) suggests a fixed ratio $L_v : L_\ell$ (e.g., 3:1 or 4:1) of the bulk phases, where we set $L_\ell = L_s$ and have chosen $L_s = 25\sigma$ or $50\sigma$ for the width of the pre-equilibrated slabs, depending on temperature.
However, due to the broadening of the interface by capillary waves, these nominal values of $L_\ell$ and $L_v$ are in general (slightly) different from the values that are deduced from the inhomogeneous sample.
For the latter step, various definitions of the mean interface position were proposed and are used in the literature
\cite{Evans:1979, Rowlinson:Capillarity, Milchev:2002, Vink:2005, Tarazona:2012, Parry:2015}.

A common choice is based on Gibbs' dividing surface (GDS), which in integral form is equivalent to
$\rho_\ell L_\ell + \rho_v L_v = N / A$;
further, $L_\ell + L_v = L_z$ is fixed to the extent of the simulation domain along the interface normal (i.e., the $z$-axis).
Combining these two relations yields $L_\ell$ for given $N$, $A$, $\rho_\ell$, $\rho_v$, and $L_z$.
On the other hand, counting the particles in the inhomogeneous system, which was assembled from pre-equilibrated bulk slabs (\cref{sec:sim_details}), yields the same expression for $N/A$ as the GDS criterion.
Hence, $L_\ell$ according to the GDS definition agrees with the nominal values for $L_\ell$.

In the present analysis, we followed a different approach and exploited the fact that there are two well-separated interfaces in the simulation setup: we have determined $L_\ell$ from fits to the simulated mean density profile $\rho_\mathrm{sim}(z)$ using an inflected sigmoidal function [\cref{eq:symmetric_sigmoidal_profile}].
The obtained values for $L_\ell$ are very close to the nominal values according to the GDS definition; the absolute deviations are less than $0.1\sigma$ at all considered temperatures, i.e., a difference of less than 4\textperthousand{}.
In addition to the values for $L_\ell$, the fits produced precise values of the coexisting densities, $\rho_\ell$ and $\rho_v$, and the interfacial width $\zeta$ (see \cref{tab:coexistence,tab:interface}), which allowed us to obtain an accurate estimate of the liquid--vapor critical point (see \cref{fig:coexistence,sec:critical}), although it is not in our focus here.

The small ambiguity in the position of the mean interface has consequences for the $q$-dependent surface tension $\gamma(q)$:
A variation of the interface position by $\delta L$ implies changing the widths of the bulk slabs from $L_\ell$ and $L_v$ to $L_\ell + \delta L$ and $L_v - \delta L$, respectively. ($\delta L$ can have either sign.)
This modifies the background contribution, given by \cref{eq:ssf_composed} in the context of the simulations, and thus the interfacial structure factor $H(q)$, which, by virtue of \cref{eq:ssf_slab_asymptotics}, receives an additive contribution
\begin{equation}
  \delta H(q) = [\rho_\ell S_\ell(q) - \rho_v S_v(q)] \, \delta L \,,
  \label{eq:delta_H_q}
\end{equation}
provided that $L_\ell \gg \xi_\ell$.
This means that $\gamma(q)$, given by \cref{eq:gamma_q_def}, is replaced by the adjusted expression
\begin{equation}
  \gamma_\mathrm{adj}(q) = \frac{\kB T (\Delta \rho)^2}{q^2[H(q) + \delta H(q)]} \,.
  \label{eq:gamma_adj}
\end{equation}
For small wave numbers, one has $\delta H(q) \ll H(q)$ [\cref{fig:gid_analysis}] so that
\begin{align}
  \gamma_\mathrm{adj}(q \to 0)
    &\simeq \gamma(q) \left[1 - \frac{q^2 \gamma(q)}{\kB T (\Delta \rho)^2} \, \delta H(q) \right] \notag \\
    &\simeq \gamma_0 \left[1+q^2(\ell^2 - \Lambda \, \delta L)\right] \,,
  \label{eq:gamma_adj_small_q}
\end{align}
where $\Lambda := \gamma_0 (\Delta \rho)^{-2} [\rho_\ell^2  \chi_T^{(\ell)}- \rho_v^2 \chi_T^{(v)}]$
is a certain length.
Inserting the values for these coefficients, as obtained in the simulations (\cref{tab:coexistence}), yields $\Lambda \approx 0.074\sigma$ for $T^*=0.70$ and
$\Lambda \approx 0.046\sigma$ for $T^*=1.15$.
With the corresponding values of $\ell$ [\cref{tab:interface,fig:ell_xi}] and assuming a physically meaningful range for $\delta L$, we conclude that the precise definition of the mean interface position has only a minor effect on the behavior of $\gamma(q)$ for small $q$.

For large wave number, however, we have $\delta H(q\to\infty) = \Delta\rho\,\delta L$. Thus, changing $H(q)$ by such an amount has the potential to qualitatively modify the behavior of $\gamma(q)$ for large $q$.
An adjustment of $L_\ell$ implies that a contribution $\propto 1/q^2$ is added reciprocally to $\gamma(q)$:
\begin{equation}
  \frac{1}{\gamma_\mathrm{adj}(q)} \simeq \frac{1}{\gamma(q)} + \frac{q^2 \delta L}{\kB T \Delta \rho} \,, \quad q \to \infty \,.
  \label{eq:gamma_adj_large_q}
\end{equation}
At large $q$, taking $\delta L < 0$ leads to an increase of $\gamma_\mathrm{adj}(q)$ relative to $\gamma(q)$
and, for $\delta L$ sufficiently large in magnitude, this can induce a bending upwards of $\gamma_\mathrm{adj}(q)$ such that $\gamma_\mathrm{adj}(q\to\infty) \to \infty$, which is desirable in the context of CW theory employing effective surface Hamiltonians \cite{Parry:1994, Mecke:1999, Tarazona:2012, Chacon:2014, Chacon:2016}.
Recently, \citet{Hernandez-Munoz:JCP2022} argued that there are no interfacial density correlations at large wave number
and they proposed to use the condition $H(q\to \infty) \to 0$ in order to tune certain parameters of the data analysis.
In our case, this amounts to adjusting the length $L_\ell$ by $\delta L = -H(q\to \infty)/\Delta\rho$, which would remove a putative $1/q^2$-contribution from $\gamma(q)$ and one would indeed obtain that $\gamma_\mathrm{adj}(q \to \infty) \to \infty$.

\begin{figure}
 \includegraphics[width=\figwidth]{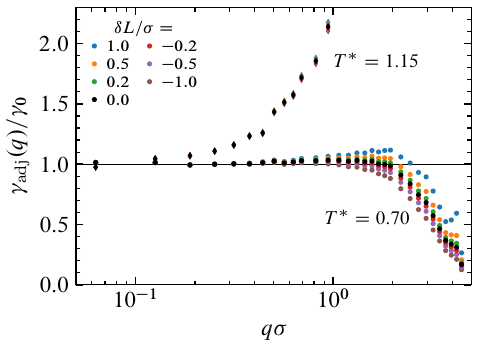}
 \caption{Robustness of the $q$-dependent surface tension with respect to a variation of the width $L_\ell$ of the liquid slab in the background contribution $S_b(q)$ by an amount $\delta L$.
 Results for this, \emph{a posteriori} adjusted quantity $\gamma_\mathrm{adj}(q)$ [\cref{eq:delta_H_q,eq:gamma_adj}],
 are shown for seven distinct values of $\delta L$ and for the temperatures $T^*=0.70$ (disks) and $T^*=1.15$ (diamonds).
 These results are based on the data provided for $\gamma(q)$ in \cref{fig:gamma_q}(a).
 }
 \label{fig:gamma_q_adjusted}
\end{figure}

For $\gamma(q)$, we have tested this procedure for the simulation results for $\gamma(q)$ shown in \cref{fig:gamma_q}(a).
The data appear to follow a decay $\gamma(q) \simeq \gamma_0 (b q)^{-2}$ for large $q$, albeit only in a small wave number window; such a decay for large $q$ would correspond to a non-zero limit of the interfacial structure factor, $H(q\to \infty) > 0$ [see \cref{eq:gamma_q_def,fig:gid_analysis}].
From fits to the data for $\gamma(q)$ in the range $2.6 \lesssim q\sigma \lesssim 3.8$, we have obtained $b = 0.41 \sigma$ and $0.21 \sigma$ at $T^*=0.70$ and $1.15$, respectively.
% T=0.70: Binf = γ₀ b^-2 = 5.1, γ₀ = 0.868, ρl = 0.824, ρv = 0.0033
% T=1.15: Binf = γ₀ b^-2 = 1.6, γ₀ = 0.069, ρv = 0.540, ρv = 0.118
The assumed asymptotic form of $\gamma(q)$ for large $q$ is equivalent to $H(q \to \infty) \simeq \kB T (\Delta\rho)^2 \, b^2 / \gamma_0 =: H_\infty$, and setting $\delta H(q\to\infty) = -H_\infty$ will remove such a spurious large-$q$ contribution from $H(q)$.
The corresponding shift $\delta L = H_\infty / \Delta \rho$ for the two temperatures renders $\delta L = 0.11 \sigma$ and $\delta L = 0.30 \sigma$, respectively.
However, shifting the mean interface position by such an amount, has only a marginal influence on $\gamma(q)$ and does not yield the desired qualitative change, namely that $\gamma(q)$ bends upwards [\cref{fig:gamma_q_adjusted}]. Instead, $\gamma(q)$, over the entire accessible range of wave numbers, is found to be rather robust against variations of $L_\ell$ within a physically meaningful range.

The reason that the above choice for $\delta L$ does not remove the apparent large-$q$ decay of $\gamma(q)$ can be understood by noting that the wave number window, within which one has $\gamma(q) \approx \gamma_0 (b q)^{-2}$, is yet to the left of the first peak of the bulk structure (which is near $q\sigma \approx 6.8$, see \cref{fig:gid_analysis}).
There, $S_\ell(q) \approx S_v(q) \approx 1$ does not hold in this regime, which was used to deduce the relation $\delta L = H_\infty / \Delta \rho$.
Therefore, in order to remove an apparent plateau in $H(q)$ around a certain intermediate wave number $q_*$, one has to consider the full $q$-dependence of $\delta H(q)$ given in \cref{eq:delta_H_q}, which suggests to set
$\delta L = -H(q_*) / [\rho_\ell S_\ell(q_*) - \rho_v S_v(q_*)]$.
At $T^*=0.70$, reasonable estimates of $S_v(q_*)$ and $S_\ell(q_*)$ are given by their values for $q\to 0$.
Using $H(q_*) = H_\infty$ one finds $\delta L = - b^2/\Lambda$ in terms of the length $\Lambda$ [introduced after \cref{eq:gamma_adj_small_q}]. This expression leads to $\delta L \approx 2.3\sigma$, which is more than four times the interfacial width $\zeta$ and thus physically unplausible.

Based on recent insight into the resonance structure of the interfacial two-point correlations,
\citet{Parry:NP2019, Parry:PRE2019, Parry:PRE2019b} have proposed that
the full wave number dependence of $\gamma_0/\gamma(q)$ is well approximated by a linear combination of the bulk structure factors $S_\ell(q) / S_\ell(q \to 0)$ and $S_v(q) / S_v(q \to 0)$ with suitable, weakly $q$-dependent weights to account for the liquid--vapor asymmetry.
Moreover, these DFT studies reveal that $H(q\to \infty) \sim q^{-2}$ (with the exception of the overly simplified square-gradient models, for which $S(q \to \infty) \sim q^{-2}$ and thus $H(q\to \infty) \sim q^{-4}$).
Concerning the present MD simulation data, we conclude that a finding of $H(q\to\infty) = H_\infty > 0$ would indeed be in conflict with the above prediction.
However, the observed decrease of $\gamma(q)$ corresponds well with the increase of $S_\ell(q)$ in the rising flank of its first peak (near $q\sigma \approx 4$, see, e.g., \cref{fig:gid_analysis}).
Moreover, the actual behavior of $\gamma(q)$ for large wave number, i.e., $q\sigma \gtrsim 5$ cannot be obtained from the data due to the unavoidable statistical noise.
Thus, from our data one cannot rule out that the actual $\gamma(q)$ has a small, positive limit $\gamma_\infty := \gamma(q \to \infty) > 0$ or, equivalently, that $H(q\to\infty) \sim q^{-2}$ --- which would be consistent with the DFT calculations.

\section{Summary and conclusions}

In sum, we have discussed the wave number-dependence of the GIXRD intensity $I(q;\kappa)$ due to scattering off liquid--vapor interfaces. We have proposed an unambiguous separation $I(q;\kappa) = I_\text{int}(q;\kappa) + I_\text{b}(q;\kappa)$  into an interface-related contribution $I_\text{int}(q;\kappa)$ and the bulk background $I_b(q;\kappa)$, as illustrated in \cref{fig:gid_analysis}; $\kappa$ is the inverse penetration length.
The separation is based on a simple reference system for the coexisting bulk phases which avoids any assumption concerning the interfacial region.
The essential ingredients are free boundary conditions for the bulk phases on both sides of the interface.
This means that  the reference system is composed of independent liquid and vapor phases and that their structures are identical to the respective bulk structures and are unperturbed by the presence of the interface.
(Necessarily, such an idealized situation cannot occur in thermal equilibrium, but only on paper, because¸ capillary waves and other interfacial fluctuations would render any physical quantity to vary smoothly across the interface.)
Accepting this simple reference model, it turns out that the background scattering $I_\ell(q;\kappa)$ from, e.g., the liquid phase is not simply proportional to the structure factor $S_\ell(q)$ of the bulk liquid [\cref{eq:gid_bulk_asymptotics}]; rather it is given as an integral over this function [\cref{eq:gid_bulk,fig:gid_bulk}].
This is a consequence of the free boundary conditions and appears likewise in the static structure factor of a liquid slab of finite width \cite{SSFslab:JCP2020} [\cref{eq:ssf_slab,eq:ssf_slab_asymptotics}].
We note that any ``continuous'' model for the background scattering, i.e., one which imposes a continuous interpolation of $G_b(q, z, z')$ across the interface, would require knowledge of the microscopic structure of the interfacial region, at least on the length scale over which the interpolation takes place.
Already at the level of the local mean density, the question how to switch between the different correlation lengths on the liquid and the vapor sides has no obvious answer without providing microscopic details.
As was shown in \cref{sec:Gb_model}, the discontinuity of the background contribution and thus of the interfacial part of the two-point correlation function $G(q, z, z')$ is compatible with the asymptotically rigorous Wertheim--Weeks result for the CW divergence [\cref{eq:WertheimWeeks}].

The interfacial part of the scattering yields the interfacial structural factor
$H(q)=I_\text{int}(q;\kappa \to 0)$
for sufficiently deep sample penetration on the liquid side ($\kappa^{-1} \gg \zeta$, which is idealized as $\kappa \to 0$). This expression of $H(q)$ defines an effective, wave number-dependent surface tension $\gamma(q)$ which is entirely based on density pair correlations [\cref{eq:gamma_q_def}].
Only for small wave numbers, $q\ell(T) \ll 1$, the classical CW divergence, i.e., $H(q) \sim q^{-2}$, is observed in the scattered intensity because in this regime $\gamma(q) \simeq \gamma_0$ [see \cref{eq:gamma_q_ell} for the definition of $\ell(T)$].
Here, we have shown that considering merely the singular part of the background contribution, $I_\ell(q;\kappa) \approx (\rho_\ell/2\kappa) S_\ell(q)$, as usually done in the analysis of GIXRD data, results in a different interfacial structure factor $\wtilde H(q)$ and, correspondingly, in a different surface tension $\tilde\gamma(q)$.
In particular, the neglected background terms do not drop out in the limit $\kappa \to 0$ but modify the surface tension at order $O(q^2)$ [\cref{eq:gamma_correction}]. The magnitude of this difference is controlled by the correction integral $\mathcal{J}_0(q)$, which is determined by the bulk structure factors [\cref{eq:correction_integral}] and which has the dimension of a length.
Depending on the temperature, $\mathcal{J}_0(q \to 0)$ can have a positive (close to the triple point temperature $T_t$) or a negative sign (close to the critical temperature $T_c$).
It turns out that at low temperatures, $\gamma(q)$ and $\tilde \gamma(q)$ exhibit qualitatively different $q$-dependences (\cref{fig:gamma_q}).
At higher temperatures, the relative difference is diminished due to the emergence of a contribution of $O(q^2)$ in $\gamma(q)$, which is characterized by a another length $\ell(T)$ that grows as $T$ is increased [\cref{eq:gamma_q_ell}].

Based on MD simulations for the truncated LJ fluid with cutoff distance $r_c=3.5\sigma$, we have presented evidence that $\ell(T)$ in fact diverges upon $T \uparrow T_c$ and that $\ell(T)$ approaches the bulk correlation lengths near criticality (\cref{fig:ell_xi}).
We observed further that the macroscopic surface tension $\gamma_0(T)$ of this truncated LJ fluid happens to be described very well by the critical scaling law along the whole coexistence line, from the triple point to the critical point (\cref{fig:gamma_crit});
the same observation was made earlier \cite{Stephan:JPCC2018,Vrabec:MP2006} for a cutoff of $r_c=2.5\sigma$.
Whereas the present study relies on LJ fluids as a generic test bed, analogous large-scale MD simulations could be performed for other substances, which would permit the direct comparison between existing GIXRD data \cite{Fradin:2000, Mora:2003, Li:2004}
and the simulation results; a similar program was carried out successfully for the bulk structure of water \cite{Clark:PNAS2010}.

At large wave number, $\gamma(q)$ is only mildly affected by the change of temperature, which together with the increase of $\ell(T)$ leads to a maximum in $\gamma(q)$ at a certain wave number.
This phenomenon was observed first in MD simulations \cite{Capillary:2015} and has since been put on firm theoretical ground by Parry \emph{et al.} \cite{Parry:2016, Parry:NP2019, Parry:PRE2019, Parry:PRE2019b}.
These theoretical studies use DFT calculations for exactly solvable models, which give insight into the structure of the two-point correlation function of the inhomogeneous fluid
and which suggests that $G(q,z,z')$ can reliably be approximated using solely the bulk structure factors and related bulk properties \cite{Parry:NP2019, Parry:PRE2019}.
The related expressions can, in principle, be translated into quantitative predictions for the scattered intensity $I(q;\kappa)$, which could be tested against its small-$\kappa$ counterpart $S_\text{tot}(q)$ in the simulations [\cref{eq:Stot-def}].
The corresponding expressions, however, are complicated by the liquid--vapor asymmetry \cite{Parry:PRE2019b} and are not yet readily available in an explicit form.
Nevertheless, our data for $\gamma(q)$, covering the full range of the wave number, are qualitatively consistent with the corresponding expectations based on the DFT approximations for $G(q,z,z')$;
this includes the possibility that $\gamma_\infty := \gamma(q \to \infty) > 0$, which cannot be resolved from the available data.
A complementary, first principles route to $G(q,z,z')$ has come into reach within a novel Barker--Henderson-like DFT treatment of inhomogeneous fluids \cite{Tschopp:PRE2020}, albeit such an endeavor may be technically challenging.

Experiments with phase-separated colloidal suspensions can, in principle, render the knowledge of the three-dimensional positions of all colloids, given the tremendous advances in confocal microscopy during the past two decades, and thus provide experimental data for $G(q,z,z')$.
In previous experiments on polymer--colloid dispersions \cite{Aarts:2004}, single scans of the focal plane perpendicular to the interface were used to obtain slices of the microscopic local density $\hat\rho(\vec r)$ (compare \cref{fig:interface2d}).
On this basis, capillary wave theory was then tested by assigning local interface positions and by calculating height-height correlation functions, closely resembling the traditional analysis of simulation data.
Yet, the reconstruction of all three-dimensional particle positions from sequences of such focal scans appears to be an ambitious task.

Here, differential dynamic microscopy (DDM) \cite{Cerbino:PRL2008,Giavazzi:PRE2009,Cerbino:JPS2021} offers an alternative: it is based on the correlation of intensity images and can yield similar information as contained in the interfacial structure factor $H(q)$ discussed in the present study.
DDM is also applicable to dense suspensions which scatter multiple times if the confocal mode of the microscope is used \cite{Lu:PRL2012}. In this case, the observation volume along the optical axis is restricted by the confocal depth, which introduces corrections in the obtained correlation functions which are analogous to the finite-$\kappa$ and finite width effects discussed here and for bulk liquids \cite{SSFslab:JCP2020}.
We expect that a refined interpretation of confocal DDM data, accounting for such corrections, can be developed along the lines presented here.

DDM is also a suitable tool for the characterization of motile suspensions, with micro-organisms or synthetic microswimmers as constituents \cite{Lu:PRL2012, Martinez:BJ2012, Kurzthaler:PRL2018, Koumakis:2018}.
Despite being inherently out of equilibrium, such suspensions exhibit a motility-induced phase separation which shares certain universal features of the liquid--vapor transition \cite{Siebert:2018,Partridge:PRL2019}.
A surface tension and a surface stiffness have been associated with simulation data for such phenomena \cite{Bialke:PRL2015,Patch:SM2018}, although a debate about even the sign of the surface tension shows that active flows and mechanical contributions must be distinguished carefully in order to arrive at a consistent physical interpretation (see Ref.~\citenum{Hermann:PRL2019} and references therein).
Similarly as for the equilibrium situation, a microscopic theory for the two-point density correlations in inhomogeneous active matter would be desirable in order to overcome the ambiguities which are associated with the notion of a fluctuating surface dividing the coexisting phases.

In a recent contribution, \citet{Hernandez-Munoz:JCP2022} discuss the predictions of extended CW theory for the surface diffraction at liquid--vapor interfaces with the fluctuating surface obtained from the intrinsic sampling method (ISM) \cite{Chacon:2003,Bresme:2008a}.
This latter approach considers the many-particle structure in the interfacial region, which is accessible within simulations, in order to define a local interface position and, in this sense, goes beyond the mere use of pair correlations as considered here.
Within both approaches to $\gamma(q)$ (i.e., via multi-particle and via pair correlations), there is consensus that the ``bending'' contribution $O(q^2)$ to the $q$-dependent surface tension should be positive; in particular, this should also hold for almost incompressible liquids at temperatures close to the triple point.
Yet, the ISM values for the corresponding length $\ell$ are considerably larger than what we have found here.
We have tested whether this difference can be diminished by tuning the slab width $L_\ell$ of the bulk liquid; $L_\ell$ enters the expression for the background contribution in the simulations [\cref{eq:ssf_composed}].
However, for the investigated LJ fluid we find that, for all accessible wave numbers, $\gamma(q)$ responds only marginally to changes of $L_\ell$ within a physically plausible range (\cref{fig:gamma_q_adjusted}).
Thus, we can conclude that our findings for $\gamma(q)$ --- as obtained from density pair correlations --- are robust with respect to the details of the definition of the mean interface position.
Moreover, the resulting shape of $\gamma(q)$ is consistent with theoretical predictions \cite{Parry:2016, Parry:NP2019, Parry:PRE2019, Parry:PRE2019b} for $G(q,z,z')$.
We note that a different, ISM-based definition of $\gamma(q)$ renders \cite{Tarazona:2012, Chacon:2014, Hernandez-Munoz:JCP2022} $\gamma(q)$ to diverge for large $q$.
We conjecture that this apparent controversy on the large-$q$ behavior of $\gamma(q)$ is a consequence of whether the definition of $\gamma(q)$ contains implicit information about three- and many-body correlations or not. This claim is motivated, first, by
noting that the ISM approach relies on this additional information whereas the present analysis of the simulation data is restricted to the use of two-point density correlations.
Second, we recall the good agreement between the data for $\gamma(q)$ as obtained along this route and the above-mentioned DFT calculations for $\gamma(q)$, both using essentially the same definition of $\gamma(q)$ in terms of $G(q,z,z')$.
Finally, we note that GIXRD experiments on fluid interfaces merely probe two-point correlations, although confining the fluid in a disordered host lattice provides scope for GIXRD studies of higher-order correlations \cite{Dietrich:1989a}.

In the context of extended CW theory, the physical interpretation of $\gamma(q)$ is broader than serving just as a proxy for the interfacial two-point correlations; rather, $\gamma(q)$ provides a mesoscopic characterization of liquid interfaces.
In this picture \cite{Napiorkowski:1993, Parry:1994, Mecke:1999, Chacon:2016}, the interface is thought of as a sharp surface which is locally dressed in an intrinsic density profile perpendicular to the surface (interpolating between the coexisting bulk phases as if there are no CWs).
The fluctuations of the local surface position are then governed by a corresponding surface Hamiltonian such that $\gamma(q) q^2$ is the free energy cost associated with surface corrugations of wave number $q$ (``capillary waves'').
Naturally, such a mesoscopic description must break down at short distances (large wave numbers), for which the molecular discreteness becomes relevant.
It has been demonstrated for solvable toy models \cite{Parry:2014} that one cannot unambiguously single out the naked CW contribution to the local density fluctuations at $O(q^2)$ due to a non-local entanglement of the two;
however, one can try to push the frontier as far as possible \cite{Tarazona:2012, Hernandez-Munoz:2016, *Hernandez-Munoz:2018, *Hernandez-Munoz:2018a}.
Overall, we can state that the long and slow-burning controversy on the concept of the wave number-dependent surface tension has been resolved, but care must be taken to respect its limitations and to not compare apples with pears.

The non-analytic contribution $O(q^2 \log(q))$ to $\gamma(q)$ due to dispersion forces is within the scope of mesoscopic surface Hamiltonians and unambiguously identifiable \cite{Napiorkowski:1993, Mecke:1999}.
This contribution results in a minimum in $\gamma(q)$ at mesoscopic wave numbers, corresponding to an enhancement of CW fluctuations. But the magnitude of this effect is not well understood yet: the minimum was found to be surprisingly shallow in simulations of untruncated LJ fluids \cite{Chacon:2014}, but sizable in experimental data from GIXRD on various liquid surfaces \cite{Fradin:2000, Mora:2003, Li:2004}.
For the latter, the correction discussed here has the potential to reduce the depth of the minimum to some extent, but we do not expect that it would qualitatively change the conclusions drawn from the experiments.
For a direct comparison, simulation data for GIXRD on liquid surfaces of other substances than LJ fluids, e.g., water, would be of great value.
It would also be of interest to highlight the role of dispersion forces in the interfacial density correlations,
exploiting recent insight into their analytic structure \cite{Parry:2016, Parry:NP2019, Parry:PRE2019, Parry:PRE2019b}.

It is straightforward to extend the concepts developed here to fluid interfaces in phase-separating binary liquid mixtures \cite{Hiester:2006,Idrissi:2015,Das:JCP2006,Roy:JCP2016,Pathania:2021}, which would lay the basis for probing local changes of the composition (and its fluctuations) in the interfacial region.
Within the Gaussian theory, the wave number-dependent surface tension $\gamma(q)$ not only determines the fluctuations of the local interface height, but also the fluctuations of the local interface normal \cite{Mecke:2005}. Therefore, in addition to GIXRD, the present results for $\gamma(q)$ are relevant to a variety of further surface-specific experimental techniques such as fluorescence spectroscopy, infrared spectroscopy and linear dichroism, generation of Maxwell displacement current (MDC), second-harmonic generation (SHG), direct measurement of the tilt angle distribution, and laser scanning confocal miscroscopy \cite{Mecke:2005}.
Furthermore, it may prove fruitful to investigate the local density correlations under non-equilibrium conditions such as
liquid--vapor interfaces in a temperature gradient \cite{Font:JPCC2018,Wilhelmsen:PRL2015,EbrahimiViand:JCP2020}.
Eventually, the wave number-dependent relaxation dynamics of capillary waves and interfacial fluctuations \cite{Zhang:L2021}
may be probed within the framework put forward here.
To this end, one merely needs to replace the static structure factors of the bulk by their corresponding intermediate scattering functions and to introduce a time lag between the factors of the two-point density correlation function [see  \cref{eq:G_microscopic}].

\begin{acknowledgments}
We thank Robert Evans, Andrew Parry, and Pedro Tarazona for helpful discussions and useful correspondence.
\end{acknowledgments}

\section*{Data availability statement}

The data that supports the findings of this study are available within the article and its supplementary material.

\appendix

\section{Molecular dynamics simulations of liquid--vapor interfaces}
\label{sec:sim_details}

Our present analytic results have been tested against large-scale MD simulations of LJ fluids with the pair potential $U(r)=4\epsilon\bigl((r/\sigma)^{-12}-(r/\sigma)^{-6}\bigr)$ truncated at the cutoff radius $r_c = 3.5\sigma$, so that $U(r > r_c) = 0$; the parameters $\epsilon$ and $\sigma$ serve as units of energy and length, respectively.
Two temperatures have been investigated in detail: $T^*=\kB T/\epsilon = 1.15$, which is close to the liquid-vapor critical point ($T_c^*\approx 1.22$), and $T^*=0.70$, slightly above the triple point temperature.
Periodic boundary conditions were applied at all faces of the cuboid simulation box with
its edge lengths chosen as $L_x = L_y < L_z$, so that stable, planar interfaces occur perpendicularly to the $z$-axis.
We used $L_x = 100\sigma$ in order to obtain a large area $A=L_x^2 = \num{e4}\sigma^2$ of the mean interface and in order to access small wave numbers $q$, which must be integer multiples of $2\pi/L_x$.
The MD simulations were carried out in the canonical ensemble with the software ``HAL's MD package'' \cite{HALMD}, which exploits the massively parallel architecture of high-end graphics processors
and which is well suited for the study of long-wavelength and low-frequency phenomena in liquids \cite{Capillary:2015,Roy:JCP2016,SSFslab:JCP2020,Straube:CP2020}.
(Concerning the relationship between the canonical and the grand canonical description of finite-size systems see Ref.~\citenum{Rohwer:PRE2019}.)
Initial and final particle configurations, time series of observables, as well as correlation functions such as $I(q;\kappa)$ were stored efficiently in the H5MD file format \cite{H5MD:2014}.
Further details on the simulations can be found in Ref.~\citenum{Capillary:2015}.

For the simulation of free liquid--vapor interfaces, a sufficiently thick film of bulk liquid is placed within the simulation domain, and the remaining space is filled with the coexisting vapor phase as to form two parallel, planar liquid--vapor interfaces (\cref{fig:snapshot-3d}).
The detailed protocol was as follows:
\begin{enumerate}[noitemsep,label={(\roman*)}]
 \item determine the coexisting liquid and vapor densities at the prescribed temperature;
 \item equilibrate the bulk phases of liquid and vapor independently, using slab-like, periodic boxes of width $L_s = 25\sigma$ ($T^*\leq 1.0$) or $50\sigma$ ($T^*\geq 1.1$);
 \item assemble the two phases, after squeezing the different configurations slightly in order to avoid particle overlaps at the boundaries between the phases (e.g., by an amount of $0.5\sigma$ along the $z$-direction via rescaling of the positions);
 \item equip the assembled system with periodic boundaries and let it relax to form an inhomogeneous fluid in equilibrium.
\end{enumerate}
For steps (i) and (ii) only relatively short simulation runs are needed, whereas much longer simulations are required for step (iv) in order to ensure the equilibration of the capillary waves, especially at small wave numbers. In step (iii) we combined several replicas of the vapor phase with one slab of liquid so that the overall box size was $L_z = 5\times 25\sigma$ at low temperatures and $L_z = 4\times 50\sigma$ at high temperatures,
yielding a total number of particles of $N=\num{209300}$ and $\num{447000}$, respectively.

After completing this procedure, a subsequent simulation is run for data production, in particular in order to calculate GIXRD intensities according to \cref{eq:gid_master}. To this end, we use the microscopic expression for the density correlation function
in terms of particle positions $\vec r_j = (\vec R_j, z_j)$:
\begin{equation}
 G(\vec q,z,z') = A^{-1} \langle \hat \rho(\vec q, z)^* \, \hat \rho(\vec q, z') \rangle ,
 \label{eq:G_microscopic}
\end{equation}
which involves the cross-sectional area $A$ and
$\hat \rho(\vec q,z) := \sum_{j=1}^N \delta(z-z_j) \exp(\i \vec q\cdot \vec R_j)$.
With this, \cref{eq:gid_master} implies
\begin{equation}
 I(|\vec q|) = A^{-1} \expect{|\hat\rho_f(\vec q)|^2}
 \label{eq:I_microscopic}
\end{equation}
in terms of the $f$-weighted density modes
\begin{equation}
 \hat\rho_f(\vec q) := \sum_{j=1}^N f(z_j - z_0) \, \exp({\i\vec q\cdot\vec R_j}) \,,
\end{equation}
where $z_0$ is the mean position of the interface and the function $f$ is given in \cref{eq:damping_finite}.
In the case $\kappa = 0$, we have $f(z)=1$ for $-L_v \leq z \leq L_\ell$. Exploiting the symmetry of the setup, we consider all $z$ values within the simulation box, i.e., we extend the computation of $\hat\rho_f(\vec q)$ to all particles.
This implies to set $L_\ell = L_s$ and to double the interfacial area such that $A$ needs to be replaced by $2A$ in expressions for the surface tension.

\section{Coexistence line and liquid--vapor critical point}
\label{sec:critical}

\begin{table*}
% define centred columns of flexible width
\renewcommand\tabularxcolumn[1]{>{\hfill}p{#1}<{\hfill\hbox{}}}
\heavyrulewidth=.1em
\small

\begin{tabularx}{\textwidth}{Xd{7}d{7}d{7}d{7}d{5}d{5}d{5}}
\toprule
% use X-columns to adjust column widths to overall text width,
% centre first part of the table rather than to align on the decimal point
\multicolumn{1}{X}{$\displaystyle \frac{\kB T}{\epsilon}$} &
\multicolumn{1}{X}{$\displaystyle \frac{P}{\epsilon\sigma^{-3}}$} &
\multicolumn{1}{X}{$\displaystyle \frac{\rho_\ell}{\sigma^{-3}}$} &
\multicolumn{1}{X}{$\displaystyle \frac{\rho_v}{\sigma^{-3}}$} &
\multicolumn{1}{X}{$\displaystyle \frac{\chi_T^{(\ell)}}{\epsilon^{-1} \sigma^3}$} &
\multicolumn{1}{X}{$\displaystyle \frac{\chi_T^{(v)}}{\epsilon^{-1} \sigma^3}$} &
\multicolumn{1}{X}{$\displaystyle \frac{\xi_\ell}{\sigma}$} &
\multicolumn{1}{X}{$\displaystyle \frac{\xi_v}{\sigma}$} \\
\midrule[\heavyrulewidth]
% temp  press  rho_ℓ  rho_v  kappa_ℓ  kappa_v  xi_ℓ  xi_v
% 0.65 &  n/a &  0.84612(5) & 0.001677(3) & 0.974(2) \\
% 0.68 &  n/a &  0.83286(5) & 0.002518(3) & 0.909(2) \\
0.70 &  0.0022(5) &  0.8239(1) & 0.0032(3) & 0.0918(2) & 471\rlap{(1)} & 0.36(2) & 0.17(5) \\
0.80 &  0.0063(2) &  0.7769(1) & 0.0085(2) & 0.1343(5) & 152\rlap{(1)} & 0.49(2) & 0.12(10) \\
0.90 &  0.0168(1) &  0.7253(1) & 0.0215(1) & 0.208(1) & 69.2(2) & 0.56(2) & 0.39(2) \\
1.00 &  0.0342(2) &  0.6658(1) & 0.0436(1) & 0.373(2) & 39.5(2) & 0.72(3) & 0.61(3) \\
1.10 &  0.0623(2) &  0.5908(3) & 0.0842(4) & 0.89(2) & 29.1(3) & 0.95(10) & 0.91(7) \\
1.15 &  0.0807(1) &  0.5403(1) & 0.1182(1) & 1.92(3) & 31.7(2) & 1.42(8) & 1.51(5) \\
\bottomrule
\end{tabularx}
\caption{
Bulk properties of truncated LJ fluids ($r_c=3.5\sigma$) along the liquid--vapor coexistence line. The densities $\rho_\ell$ and $\rho_v$ of the coexisting liquid and vapor phases were determined from the density profiles obtained in simulations of the inhomogeneous system. The corresponding pressures $P$, the isothermal compressibilities $\chi_T^{(\ell)}$ and $\chi_T^{(v)}$, and the correlation lengths $\xi_\ell$ and $\xi_v$ stem from separate simulations of the bulk phases;
the last four quantities were calculated from OZ fits to the bulk structure factors [\cref{eq:ornstein-zernike}].
The numbers in parentheses give the measurement uncertainty in the last digit.
}
\label{tab:coexistence}
\end{table*}

% pressure from low-density virial equation of state
% (Hansen/McDonald, eq. 3.9.10):
%
% T = 0.70,  ρ = 0.003235, P = 0.00219
% T = 0.80,  ρ = 0.009357, P = 0.00695
% T = 0.90,  ρ = 0.02157,  P = 0.0168
% T = 1.00,  ρ = 0.04364,  P = 0.0339
% T = 1.10,  ρ = 0.0842,   P = 0.0589

The simulation setup contains two well-separated and independent interfaces (\cref{fig:snapshot-3d}). For each investigated temperature, we have fitted the simulated mean density profile $\rho_\mathrm{sim}(z)$ with an inflected sigmoidal function:
\begin{equation}
  \rho_\mathrm{sim}(z) =
  \frac{\rho_\ell + \rho_v}{2} -  \frac{\Delta \rho}{2} \, \tanh\mleft(\frac{|z-z_m| - L_s / 2}{2 \zeta}\mright) \,;
  \label{eq:symmetric_sigmoidal_profile}
\end{equation}
$z_m$ denotes the symmetry center of the liquid slab and belongs to the set of fit parameters.
This yields precise estimates of the width $L_s$ of the liquid slab (and thus of the mean interface positions $z_m \pm L_s / 2$), but also of the coexisting number densities $\rho_\ell$ and $\rho_v$, and of the interfacial width $\zeta$, which are reported in \cref{tab:coexistence,tab:interface}.
The results for $\rho_\ell$ and $\rho_v$ indicate the position of the binodal curve of the liquid--vapor transition in the temperature--density plane [\cref{fig:coexistence}(a)].
Anticipating the critical scaling behavior of the density difference, i.e., $\Delta\rho \sim (T_c-T)^\beta$ upon $T \uparrow T_c$, we estimated the critical temperature $T_c=(1.215\pm0.001) \epsilon/\kB$ from a linear regression to the rectified data [\cref{fig:coexistence}(b)];
this value is in good agreement with earlier simulation data \cite{Dunikov:2001}.
In contrast to Ising spin models, the liquid--vapor binodal is asymmetric. However, the mean density $\rho_\mathrm{sym}(T) = [\rho_\ell(T) + \rho_v(T)] / 2$ serves as a symmetry line of the binodal, which is found to be almost a straight line [\cref{fig:coexistence}(a)].

The latter observation is phenomenologically known as the ``law of rectilinear diameter'' \cite{Zollweg:1972,Wilding:1995}.
From the linear extrapolation of $\rho_\mathrm{sym}(T)$ to $T_c$, the critical density was found to be $\rho_c = (0.318 \pm 0.001) \sigma^{-3}$.
In a more refined analysis of the critial behavior of the coexisting densities, it is argued that the slope of the curve $\rho_\mathrm{sym}(T)$ vs. $T$ is proportional to the isochoric specific heat $c_V$, so that one expects a singular dependence \cite{Zollweg:1972,Sengers:1986,Wilding:1997}:
$\rho_\mathrm{sym}(T) - \rho_c \sim (T_c - T)^{1-\alpha}$ as $T \uparrow T_c$,
where $\alpha\approx 0.110$ is the Ising universal exponent of the specific heat.
However, only the two data points for $\rho_\mathrm{sym}(T)$ which are closest to $T_c$ (i.e., for $T^* \geq 1.10$) are compatible with this singular scaling law --- the linear law (i.e., mean-field like with $\alpha=0$) provides a better description of the data.
A similar observation was made before for other simple fluids (see, e.g., Refs.~\citenum{Zollweg:1972} and \citenum{Wilding:1995});
yet, it is particularly surprising here, given that the $T$-dependences of $\Delta \rho$ and $\gamma_0$ were very well captured by their corresponding critical laws over a wide range of temperatures (\cref{fig:coexistence}(b) and \cref{fig:gamma_crit}).
These findings underscore that the true scaling behavior sets in only asymptotically for $T \uparrow T_c$.
In particular, the temperature dependence of $c_V$ is found to be non-monotonic along the liquid branch of the coexistence line and has its minimum near $T^* \approx 1.05$ (data not shown; this calls for future research).

Along the transition line, we have also computed the pressures of the coexisting liquid and vapor phases from the bulk simulations [\cref{tab:coexistence} and \cref{fig:coexistence}(c)], which served as a consistency check.
Eventually, the critical pressure was obtained from a separate simulation of the bulk fluid at the quoted critical point $(T_c, \rho_c)$, yielding $P_c = (0.11074 \pm 0.00003) \epsilon \sigma^{-3}$.

\bibliography{capillary}

\end{document}